\documentclass[12pt,english]{article}

\usepackage{mathptmx}
\usepackage[T1]{fontenc}
\usepackage[latin9]{inputenc}
\usepackage{geometry}
\usepackage{color}
\usepackage{babel}
\usepackage{mathtools}
\usepackage{amsmath}
\usepackage{amssymb}

\usepackage[authormarkup=none,draft]{changes}
\definechangesauthor[name={del}, color=red]{del}

\usepackage{stmaryrd}
\SetSymbolFont{stmry}{bold}{U}{stmry}{m}{n}
\usepackage{graphicx}
\usepackage{setspace}
\usepackage[unicode=true,pdfusetitle,
 bookmarks=true,bookmarksnumbered=false,bookmarksopen=false,
 breaklinks=false,pdfborder={0 0 1},backref=false,colorlinks=true]
 {hyperref}
\hypersetup{
 linkcolor=blue,urlcolor={blue},filecolor={blue},citecolor={blue}}

\makeatletter
\newcommand{\lyxaddress}[1]{
	\par {\raggedright #1
	\vspace{1.4em}
	\noindent\par}
}

\usepackage{chngcntr}
\usepackage{mathtools}
\counterwithout{figure}{section}
\counterwithout{equation}{section}
\usepackage{times}
\usepackage{amsmath}
\usepackage{floatrow}	

\usepackage{times}
\usepackage{upgreek}

\usepackage[caption=false]{subfig}
\usepackage{floatrow}

\counterwithout{figure}{section}
\counterwithout{equation}{section}
\usepackage{changepage}

\usepackage{babel}

\counterwithout{figure}{section}
\counterwithout{equation}{section}
\usepackage{times}
\usepackage{float}

\usepackage{bm}
\usepackage[caption=false]{subfig}

\usepackage[numbers,sort&compress]{natbib} 

\date{}

\renewcommand{\fnum@figure}{Fig. \thefigure}

\makeatother

\begin{document}
\global\long\def\dispv{\mathbf{u}}%
\global\long\def\forcev{\mathbf{f}}%
\global\long\def\stresst{\boldsymbol{\sigma}}%
\global\long\def\grad{\nabla}%
\global\long\def\div{\nabla\cdot}%
\global\long\def\curl{\nabla\times}%

\global\long\def\trans{t}%
\global\long\def\refr{r_{R}}%
\global\long\def\refl{r_{L}}%
\global\long\def\kiny{k_{1}}%
\global\long\def\kinyd{\hat{k}_{1}}%
\global\long\def\linni{\hat{k}_{i}}%
\global\long\def\kmb{\delta}%
\global\long\def\slown{\kmb}%
\global\long\def\ufx{Z}%
\global\long\def\ufy{X}%
\global\long\def\dxo{\partial_{(1)}}%
\global\long\def\dxt{\partial_{(2)}}%
\global\long\def\dxi{\partial_{(i)}}%
\global\long\def\ntnf{L_{1}}%
\global\long\def\ntns{L_{2}}%
\global\long\def\ntni{L_{i}}%
\global\long\def\rhom{\rho_{p}}%
\global\long\def\ksh{k_{h}}%
\global\long\def\rhoz{\rho_{0}}%
\global\long\def\linnf{\frac{\rhom}{\rhoz}\frac{\ksh^{2}}{\kinx}}%
\global\long\def\btl{p}%
\global\long\def\sta{s}%
\global\long\def\solf{\psi_{+}}%
\global\long\def\solb{\psi_{-}}%
\global\long\def\psm{\mathsf{Q}}%
\global\long\def\trm{\mathsf{M}}%
\global\long\def\fwro{w_{\nu}}%
\global\long\def\xdt{\ufx_{,z}}%
\global\long\def\tranmb{\hat{\text{U}}}%
\global\long\def\nvar{\zeta}%
\global\long\def\xls{0}%
\global\long\def\xr{L}%
\global\long\def\poy{P}%
\global\long\def\besf{C_{1}}%
\global\long\def\besb{C_{2}}%
\global\long\def\paw{\alpha}%
\global\long\def\Ef{\ufx}%
\global\long\def\Efone{\Ef^{(1)}}%
\global\long\def\Eftwo{\Ef^{(2)}}%
\global\long\def\Efi{\Ef^{(i)}}%
\global\long\def\epf{A_{f}}%
\global\long\def\epb{B_{b}}%
\global\long\def\Eff{\Ef_{f}}%
\global\long\def\Efb{\Ef_{b}}%
\global\long\def\pmpsi{\psi_{\pm}}%
\global\long\def\eigsc{\lambda}%
\global\long\def\wavelz{c_{0}}%
\global\long\def\eigvsc{v}%

\global\long\def\pvity{\epsilon}%
\global\long\def\pbili{\mu}%
\global\long\def\magf{B}%
\global\long\def\elf{E}%
\global\long\def\elfv{\boldsymbol{E}}%
\global\long\def\magfv{\boldsymbol{B}}%
\global\long\def\eztep{\xi}%
\global\long\def\ind{n}%
\global\long\def\epsa{\tilde{\eztep}}%
\global\long\def\xx{x_{3}}%
\global\long\def\xy{x_{1}}%
\global\long\def\xz{x_{2}}%
\global\long\def\dz{_{,2}}%
\global\long\def\dx{_{,3}}%
\global\long\def\dy{_{,1}}%
\global\long\def\ddz{_{,22}}%
\global\long\def\ddx{_{,33}}%
\global\long\def\ddy{_{,11}}%
\global\long\def\xl{z}%
\global\long\def\pvityv{\epsilon_{0}}%
\global\long\def\pbiliv{\mu_{0}}%
\global\long\def\pvityh{\epsilon_{h}}%
\global\long\def\pvityp{\epsilon_{p}}%
\global\long\def\uamf{\elf_{f}}%
\global\long\def\uamb{\elf_{b}}%
\global\long\def\trmc{M}%
 
\global\long\def\xr{L}%
\global\long\def\xp{y}%
\global\long\def\xt{x}%
\global\long\def\freq{\omega}%
\global\long\def\px#1{\partial_{#1}}%
\global\long\def\light{c}%
\global\long\def\il{A_{1}}%
\global\long\def\ol{B_{1}}%
\global\long\def\ir{A_{2}}%
\global\long\def\orr{B_{2}}%
\global\long\def\order{\nu}%
\global\long\def\swave{\text{s\ wave}}%
\global\long\def\grating{\beta}%
\global\long\def\fb{\frac{\pi}{\xr}}%
\global\long\def\coft{\sqrt{\frac{\ksh^{2}\pvityp}{-\grating^{2}\pvityh}}}%
\global\long\def\st{\tilde{\mathsf{S}}}%
\global\long\def\sr{\mathsf{S}}%
\global\long\def\phaz{z_{0}}%
\global\long\def\pt{\mathcal{PT}}%
\global\long\def\disy{u}%
\global\long\def\pop{\mathcal{P}}%
\global\long\def\top{\mathcal{T}}%
\global\long\def\scaxo{\xl^{(1)}}%
\global\long\def\scaxt{\xl^{(2)}}%
\global\long\def\scaxi{\xl^{(i)}}%
\global\long\def\scaxth{\xl^{(3)}}%
\global\long\def\paulies#1{\sigma_{#1}}%

\global\long\def\besn{\nu}%
\global\long\def\lcof{\mathsf{l}}%
\global\long\def\rcof{\mathsf{r}}%
\global\long\def\cvec{\mathsf{c}}%
\global\long\def\qhom{\mathsf{Q}_{h}}%
\global\long\def\mcomp#1{M_{#1}}%
\global\long\def\scomp#1{S_{#1}}%
\global\long\def\evs{\lambda^{\left(\sr\right)}}%
\global\long\def\evsc{\lambda^{\left(\st\right)}}%
\global\long\def\Tran{T}%
\global\long\def\RL{R_{L}}%
\global\long\def\RR{R_{R}}%
\global\long\def\ev{\lambda}%
\global\long\def\ang{\theta}%
\global\long\def\RLorR{R_{L/R}}%
\global\long\def\rLorR{r_{L/R}}%
\global\long\def\fpoles{f}%
\global\long\def\fzeroes{g}%
\global\long\def\evsnum#1{\evs_{#1}}%
\global\long\def\phazpol#1{\phaz^{\left(#1\right)}}%
\global\long\def\polsind{\nu}%
\global\long\def\angpol#1{\ang^{\left(#1\right)}}%
\global\long\def\phazer#1{\widetilde{\phaz}^{\left(#1\right)}}%
\global\long\def\angzer#1{\widetilde{\ang}^{\left(#1\right)}}%
\global\long\def\chind{n}%
\global\long\def\chindr{n'}%
\global\long\def\chindi{n''}%
\global\long\def\freqp#1{^{\left(#1\right)}}%
\global\long\def\freqz#1{\widetilde{\freq}^{\left(#1\right)}}%
\global\long\def\freqr{\freq_{r}}%
\global\long\def\freqi{\freq_{i}}%
\global\long\def\phazsol#1{^{#1}}%
\global\long\def\ptop#1{\tilde{\bar{#1}}}%
\global\long\def\kinx{\ksh\sin\ang}%
\global\long\def\nn{\nu_{0}}%
\global\long\def\nnb{m}%
\global\long\def\bessel{J}%
\global\long\def\besselt{J'}%
\global\long\def\xiz{C}%
\global\long\def\nnbs{n}%
\global\long\def\besseli{I}%
\global\long\def\state{\mathsf{s}}%

\title{Oblique scattering from non-Hermitian optical waveguides}
\author{Tal Goldstein and Gal Shmuel}
\maketitle

\lyxaddress{\begin{center}
Faculty of Mechanical Engineering, Technion--Israel Institute of
Technology, Haifa 32000, Israel
\par\end{center}}
\begin{abstract}
A judicious design of gain and loss leads to counterintuitive wave
phenomena that are inaccessible by conservative systems. Notably,
such designs can give rise to laser-absorber modes and anisotropic
transmission resonances. Here, we analyze the emergence of these phenomena
in an optical scatterer with sinusoid gain-loss modulation that is
subjected to monochromatic oblique waves. We derive an analytical
solution to the problem, with which we show how the scatterer parameters,
and specifically the modulation phase and incident angle, constitute
a real design space for these phenomena. 
\end{abstract}

\section{Introduction\label{sec:Introduction}}

A judicious design of gain and loss in physical systems in general
\cite{El-Ganainy2018ys,Ashida2020nonhermitian}, and in particular
optical systems \cite{Feng2017np,El-Ganainy2019cp,Midya2018natcom,Ozdemir2019cr,Zhao2018nsr,Pilehvar2022optexpress},
leads to counterintuitive wave phenomena that are inaccessible by
conservative systems. These phenomena, such as supersensitivity \cite{Wiersig2014prl,Zhong2019prl,Djorwe2019prapplied}
and unidirectional invisibility \cite{lin2011prl,Mostafazadeh2013pra},
can be harnessed for different engineering applications that require
wave manipulation. 

The bulk of the works is on systems that are invariant to combined
parity-time ($\pt$) transformations \cite{Bender1998PRL,christodoulides2018parity,Fleury2015,Graefe2011pra,Hou2018PRApplied,ruter2010observation,Christensen2016prl},
which translates to the condition $\pvity\left(-\xl\right)=\pvity^{*}\left(\xl\right)$
for the dielectric coefficient of a medium that
is modulated along $\xl$. Depending on the parameters of such systems,
their spectrum can be real, in spite of the fact that they are governed 
by non-Hermitian operators \cite{Mostafazadeh2002jmp,moiseyev2011book}.
This domain in the parameter space is called the $\pt$-exact phase,
and the rest of the domain, at which the eigenvalues are complex,
is called the $\pt$-broken phase. The transition points between the
two phases are a type of \emph{exceptional points} (EPs) \cite{berry2004,Heiss2012jpa,Heiss2016yt,Mailybaev2005PRA,Miri2019science},
at which the eigenvalues and eigenvectors become degenerate, and are
the source of the counterintuitive phenomena mentioned earlier \cite{Achilleos2017PRB,elbaz2022jphysd,geng2021prsa,Goldzak2018PRL,hodaei2017enhanced,Longhi2018optlett,Lustig2019,Peng2016pnas,pick2017prb,Shen2018prmat,shi2016accessing,Sweeney2019PRL,Yin2013rq}. 

Of particular relevance to this work is the phenomenon of unidirectional
invisibility, which was discovered by \citet{lin2011prl}, when analyzing
a $\pt$-symmetric scatterer with sinusoid modulation.  Using the
rotating wave approximation, they have found that at the EPs of the
scattering matrix, the reflection vanishes from one side only, while
the transmission is unity; they termed it later as anisotropic transmission
resonance (ATR) \cite{ge2012conservation}. In addition, the transmission also has zero phase,
hence the scatterer is unidirectionally invisible. Later on, \citet{Longhi2011},
\citet{Jones2012} and \citet{uzdin2012} used exact solutions to
analyze the scattering properties beyond the limitations of the rotating
wave approximation. 

A second pioneering work that motivated the study to follow is by
\citet{Longhi2010pra}. He showed that a waveguide with uniform grating
and two symmetric layers of gain and loss can simultaneously act as
a laser oscillator, emitting coherent waves \cite{Siegman1986eu},
and as a coherent perfect absorber (CPA), completely absorbing particular
incoming waves \cite{Chong2010PRL}. \citet{chong2011p} have identified
the CPA-laser states as special solutions in the $\pt$-broken phase,
where a pole and a zero of the scattering matrix\footnote{This scattering matrix is different than the scattering matrix whose
EPs correspond to unidirectional invisibility, as detailed later in
Sec.\ \ref{sec:Scattering-analysis}. } coincide.

The pioneering works in Refs.\ \cite{Longhi2010pra,lin2011prl} led
to various studies whose objective is to control the phase transition
and scattering singularities by different means, such as the incident
angle of oblique waves and the chirality in a single gain-loss bilayer
\cite{Droulias2019prb,katsantonis2020pt}; or the sinusoid modulation
properties \cite{liu2020scattering}. Here, we extend the study 
to the problem of oblique waves that are scattered by waveguides of
different sinusoid gain-loss modulations. We derive an exact analytical
solution to the problem, with which we characterize how the scattering
properties, and specifically the EPs, depend on the parameters of
the system, such as the incident angle ($\ang$); driving frequency
($\freq$); and the wavelength- ($\grating$), amplitude- and phase
($\phaz$) of the modulation. For two particular cases where the modulation
yields a $\pt$-symmetric medium, we characterize the $\pt$ phase
diagram in the $\left(\ang,\freq\right)$ and $\left(\ang,\grating\right)$
parameter spaces. We also calculate the phase diagram that defines
the ATRs in the $\left(\ang,\grating\right)$ space. We show that
there is a range of modulation amplitudes and wavelengths at which
these ATRs coincide with (or reside very close to) Fabry-P\'erot resonance frequencies.
By analyzing the resultant structure of the phase diagram, we also
gain insights on how to access bidirectional zero-reflection states.
We conclude the study with an analysis of the poles and zeros of the
scattering matrix. We show that the  modulation phase and incident
angle constitute together real (rather than complex) design space
for these singularities, and specifically for quasi CPA-laser states.

Our results are presented in the following order. Sec.\ \ref{sec:Problem-statement-and}
contains the mathematical formulation of the problem, together with
our derivation of its analytical solution. In Sec.\ \ref{sec:Scattering-analysis},
we recall the two definitions of the scattering matrix and their connection
to EPs. We further derive a useful relation between the poles and
zeros of the two matrix definitions, for the particular family of
modulations that we analyze. We carry out a parametric study in Sec.\ \ref{sec:Examples},
and conclude this paper with a summary of our main results in Sec.\ \ref{sec:Summary}.

\section{Problem statement and exact solution\label{sec:Problem-statement-and}}

\floatsetup[figure]{style=plain,subcapbesideposition=top}

\begin{figure}
\centering\sidesubfloat[]{{\small{}\label{fig:IncidentAndOutcoming}}\includegraphics[scale=0.4]{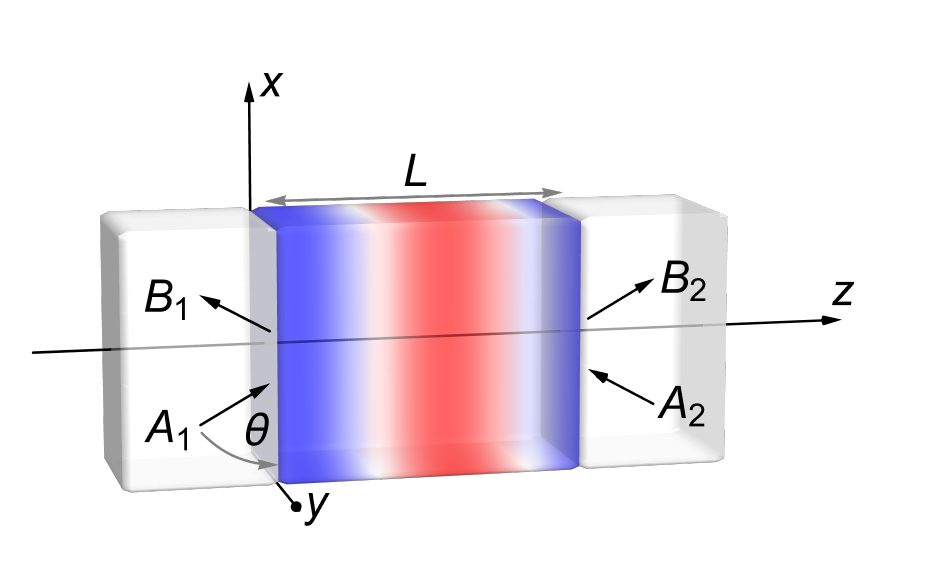}} \quad \centering\sidesubfloat[]{{\small{}\label{fig:Modulation}}\includegraphics[scale=0.4]{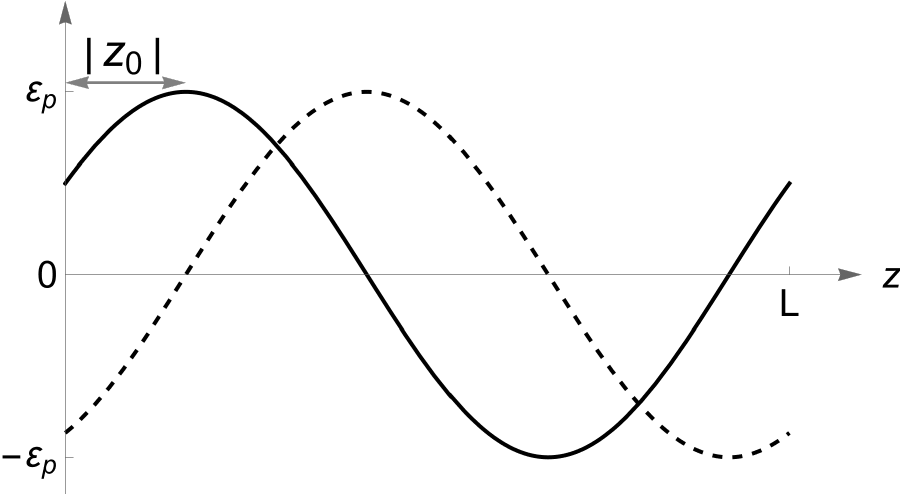}}

\caption{(a) A nonmagnetic modulated waveguide that is connected to two uniform
waveguides at $\protect\xl=0$ and $\protect\xr$. The coefficients
$\protect\il$ and $\protect\ir$ ($\protect\ol$ and $\protect\orr$)
are identified with the amplitudes of the incoming (outgoing) waves.
The dielectric coefficient of the modulated waveguide is $\protect\pvity\left(\protect\xl\right)=\protect\pvityh+\protect\pvityp e^{i2\protect\grating\left(\protect\xl+\protect\phaz\right)}$.
(b) An exemplary modulation profile for some $\protect\phaz$. The
continuous and dashed curves correspond to $\text{Re}\,\protect\pvity$
and $\text{Im}\,\protect\pvity$, respectively. The latter is also
illustrated in the previous panel using the blue-red color map, designating
gain and loss, respectively. }

{\small{}{}\label{fig:system}}{\small\par}
\end{figure}
We consider a nonmagnetic medium at $\xls\leq\xl\leq\xr$, whose dielectric
coefficient is modulated according to 
\begin{equation}
\pvity\left(\xl\right)=\pvityh+\pvityp e^{i2\grating\left(\xl+\phaz\right)}.\label{eq:modulation}
\end{equation}
where $\grating=\pi/\xr$ is half the wavenumber of the modulation.
Note that when $\phaz=0$ or $\xr/2$, the dielectric coefficient
satisfy the necessary condition for $\pt$ symmetry, namely, $\pvity\left(-\xl\right)=\pvity^{*}\left(\xl\right)$.
(The particular case of $\phaz=0$ was considered in Refs.\ \cite{lin2011prl,Jones2012,Longhi2011,uzdin2012},
for normal incident waves.) Two homogeneous nonmagnetic waveguides
with a dielectric constant $\pvityh$ are connected at $\xl\leq\xls$
and $z\geq\xr$, guiding oblique monochromatic waves to- and from
the modulated medium (Fig.\ \ref{fig:system}). We focus on TE (or
$H$) modes \cite{Tamir1964ieee}, such that the electric field is
in the $\xp$ direction, and propagates in the $x\xl$ plane. The
governing equation for these modes is 

\begin{equation}
\left[\px{\xt}^{2}+\px{\xl}^{2}+\frac{\freq^{2}}{\light^{2}}\pvity\left(\xl\right)\right]\elf\left(\xt,\xl\right)=0,\label{eq:eqvito2}
\end{equation}
where $\light$ is the speed of light, and we consider time dependency
of the form $e^{-i\freq t}$ with an angular frequency $\freq$. The
solution in the homogeneous waveguides is given by 
\begin{equation}
\elf\left(\xt,\xl\right)=\begin{cases}
\il e^{i\ksh\left(\xt\cos\theta+\xl\sin\theta\right)}+\ol e^{i\ksh\left(\xt\cos\theta-\xl\sin\theta\right)} & \xl\leq\xls,\\
\ir e^{i\ksh\left[x\cos\theta-\left(\xl-\xr\right)\sin\theta\right]}+\orr e^{i\ksh\left[\xt\cos\theta+\left(\xl-\xr\right)\sin\theta\right]} & z\geq\xr,
\end{cases}\label{eq:homogeneous fields}
\end{equation}
where $\il$ and $\ir$ ($\ol$ and $\orr$) are the amplitudes of
the incoming (outgoing) waves, $\ksh=\freq\sqrt{\pvityh}/\light$,
and $\theta$ is the angle between the $\xt\xp$ plane and the waves
in these waveguides. 

Inside the modulated medium, we seek solutions in the form 
\begin{equation}
\elf\left(\xt,\xl\right)=e^{i\ksh x\cos\theta}\Ef\left(\xl\right),\label{eq:anzats}
\end{equation}
where the $\xt$ dependency is enforced by the continuity of $\elf$
at $\xl=\xls$ and $\xr$. By substituting the ansatz \eqref{eq:anzats}
into Eq.\ \eqref{eq:eqvito2}, we obtain the following equation for
$\Ef\left(\xl\right)$ 
\begin{equation}
\left[\px{\xl}^{2}+\frac{\freq^{2}}{\light^{2}}\left(\pvityh+\pvityp e^{i2\grating\left(\xl+\phaz\right)}\right)-\ksh^{2}\cos^{2}\theta\right]\Ef\left(\xl\right)=0,\label{eq:zeq}
\end{equation}
which can rearranged as 
\begin{equation}
\left[\px{\xl}^{2}+\ksh^{2}\sin^{2}\theta\left(1+\frac{1}{\sin^{2}\theta}\frac{\pvityp}{\pvityh}e^{i2\grating\left(\xl+\phaz\right)}\right)\right]\Ef\left(\xl\right)=0.\label{eq:zeq2}
\end{equation}
Our next step is to rewrite Eq.\ \eqref{eq:zeq2} as a Bessel equation. To this end, we perform a change of variable, namely,
\begin{equation}
\nvar=\coft e^{i\grating\left(\xl+\phaz\right)},\label{eq:change of variable}
\end{equation}
such that
\begin{equation}
\px{\xl}\nvar=i\beta\nvar,\quad\px{\xl}^2\nvar=-\beta^2\nvar,\label{eq:derivatives}
\end{equation}
which together with the chain rule $\px{\xl}\Ef=\partial_\nvar Z\partial_z\nvar$ allows us to replace  Eq.\,\eqref{eq:zeq2} with

\begin{equation}
\left[\nvar^{2}\px{\nvar}^{2}+\nvar\px{\nvar}+\left(\nvar^{2}-\frac{\ksh^{2}}{\grating^{2}}\sin^{2}\theta\right)\right]\Ef=0.\label{eq:bessel}
\end{equation}
Eq.\ \eqref{eq:bessel} is solved exactly using the Bessel functions
$\bessel_{\pm\order}\left(\nvar\right)\eqqcolon\pmpsi\left(\xl\right)$,
where $\order=\left(\ksh/\grating\right)\sin\theta$ is assumed to
be a non-integer number. Essentially, Eq.\ \eqref{eq:bessel} and
its solution are generalizations of the results in Refs.\ \cite{Longhi2011,Jones2012,uzdin2012}
to oblique waves.  The electric field in the modulated medium is
a linear combination of these solutions, such that 
\begin{equation}
\elf\left(\xt,0\leq\xl\leq\xr\right)=\besf e^{i\ksh x\cos\theta}\solf\left(\xl\right)+\besb e^{i\ksh\xt\cos\theta}\solb\left(\xl\right).\label{eq:sol medium}
\end{equation}
Our next objective is to related the amplitudes of the waves inside the modulated medium to the amplitudes in the homogeneous waveguides. To this end, we assemble the two quantities that are continuous along $z$, i.e.,  $\elf$ and $\px{\xl}\elf$, into a column vector which we denote by $\state$. In the modulated medium, this so-called state vector can be written as
\begin{equation}
\state\left(0\leq\xl\leq\xr\right)=\mathsf{Q}\left(z\right)\mathsf{c},
\label{eq:statecenter}
\end{equation}
where
\begin{align*}
\text{\ensuremath{\psm}}\left(\xl\right)=\left(\begin{array}{cc}
\solf\left(\xl\right) & \solb\left(\xl\right)\\
\px{\xl}\solf\left(\xl\right) & \px{\xl}\solb\left(\xl\right)
\end{array}\right), \cvec=\left(\begin{array}{c}
\besf\\
\besb
\end{array}\right).
\end{align*}

Similarly, the state vector in the right- and left ends of the left- and right homogeneous waveguides is
\begin{equation}
\state\left(0^-\right)=\mathsf{Q}_h\mathsf{l},\quad\state\left(L^+\right)=\mathsf{Q}_h\mathsf{r},
\label{eq:stateends}
\end{equation}
respectively, where
\begin{align*}
\qhom & =\left(\begin{array}{cc}
1 & 1\\
i\ksh\sin\theta & -i\ksh\sin\theta
\end{array}\right),
\lcof=\left(\begin{array}{c}
\il\\
\ol
\end{array}\right), \rcof=\left(\begin{array}{c}
\ir\\
\orr
\end{array}\right).
\end{align*}
It now follows that
from the continuity of the state vector  at $\xl=\xls$
and $\xr$, we obtain 
\begin{equation}
\qhom\lcof=\text{\text{\ensuremath{\psm}}\ensuremath{\left(0\right)}}\cvec,\ \qhom\rcof=\text{\text{\ensuremath{\psm}}\ensuremath{\left(\xr\right)}}\cvec.\label{eq:boundary continouity}
\end{equation}
Accordingly, the transfer matrix $\trm$  that relates between the
wave amplitudes of the two homogeneous waveguides is 
\begin{equation}
\trm=\qhom^{-1}\text{\ensuremath{\psm}}\left(\xr\right)\text{\ensuremath{\psm}}^{-1}\left(0\right)\qhom,\ \text{such\ that}\ \rcof=\trm\lcof;\label{eq:transfer mat}
\end{equation}
this completes the solution for the electric waves, given any amplitude
of incoming waves. Using the Bessel's continuation rule we have that
$\text{\ensuremath{\psm}}\left(\xr\right)=\text{\ensuremath{\psm}}\left(0\right)\left(\begin{array}{cc}
e^{i\ksh\sin\ang\xr} & 0\\
0 & e^{-i\ksh\sin\ang\xr}
\end{array}\right)$, which together with Eq.\ \eqref{eq:transfer mat} yields $\det\trm=1$.

\section{Scattering analysis\label{sec:Scattering-analysis}}

We can now relate the incoming and outgoing waves in terms of the
components of $\trm$. One way to do that is 
\begin{equation}
\left(\begin{array}{c}
\ol\\
\orr
\end{array}\right)=\sr\left(\begin{array}{c}
\il\\
\ir
\end{array}\right),\ \sr=\left(\begin{array}{cc}
\mcomp{12}/\mcomp{22} & \mcomp{22}^{-1}\\
\mcomp{22}^{-1} & -\mcomp{21}/\mcomp{22}
\end{array}\right).\label{eq:sdef}
\end{equation}
If we prescribe an incoming wave with unitary amplitude from the left
(right), such that $\il=1$ and $\ir=0$ ($\il=0$ and $\ir=1$),
it follows that the amplitude of the transmitted wave is $\scomp{21}$
($\scomp{12}$), and the amplitude of the reflected wave is $\scomp{11}$
($\scomp{22}$). Accordingly, the components of $\sr$ are identified
with the reflection and transmission coefficients, such that 
\begin{equation}
\sr=\begin{pmatrix}\refl & t\\
t & \refr
\end{pmatrix}.\label{eq:sdef coefficients}
\end{equation}
The eigenvalues of $\sr$ are 
\begin{equation}
\evs_{1,2}=\frac{\refl+\refr}{2}\pm\sqrt{\frac{\left(\refl-\refr\right)^{2}}{4}+t^{2}};\label{eq:sev}
\end{equation}
they become degenerate, together with their eigenvectors, when $\left(\refl-\refr\right)/t=\pm2i$.
These exceptional points of $\sr$ for a $\pt$-symmetric scatterer
are linked to the degeneracies in its eigenmodes, if it was bounded
\cite{Ambichl2013prx}. 

There is an alternative scattering matrix, denoted here $\st$, that is obtained if we interchange the two entries of the column vector in the left-hand-side of Eq.\,\eqref{eq:sdef}. By doing so, we obtain
\begin{equation}
\left(\begin{array}{c}
\orr\\
\ol
\end{array}\right)=\st\left(\begin{array}{c}
\il\\
\ir
\end{array}\right),\ \st=\begin{pmatrix}t & \refr\\
\refl & t
\end{pmatrix}.\label{eq:alternative s}
\end{equation}
Both Eq.\,,\eqref{eq:sdef} and Eq.\,\eqref{eq:alternative s} deliver the same relations between the incoming and outgoing waves, however the corresponding scattering matrices have different eigenvalues. Specifically, the eigenvalues of $\st$ are 
\begin{equation}
\evsc_{1,2}=t\pm\sqrt{\refl\refr};\label{eq:sev-2}
\end{equation}
they become degenerate when either $\refl$ or $\refr$ vanish. When the scatterer is $\pt$-symmetric,
the exceptional points of $\st$ reflect anisotropic transmission
resonances (ATRs), at which the reflection may vanish only from one
side while the transmittance is unity, since the scattering coefficients
for $\pt$-symmetric system satisfy
\begin{equation}
\left|\Tran-1\right|=\sqrt{\RL\RR},\label{eq:conservation}
\end{equation}
where $\Tran\coloneqq\left|t\right|^{2}$ is the transmittance, and
$R_{L,R}\coloneqq\left|r_{L,R}\right|^{2}$ are the two reflectances.

The zeroes of $\sr$, which correspond to a zero eigenvalue, reflect
a perfectly absorbing medium, while the poles of $\sr$, which correspond
to an infinite eigenvalue, reflect a lasing oscillator which emits outgoing
coherent waves. It is follows that if $\sr$ has a zero, then from
Eq.\ \eqref{eq:sev} we have $t^{2}=\refl\refr$, thus $\st$ has
a zero too. In our case, it also follows that if $\sr$ has a pole
then $\st$ has a pole too. To show this, we note that
\begin{align}
\paulies 1\sr^{*}\left(-\phaz\right)\paulies 1 & =\sr^{-1}\left(\phaz\right),\st^{*}\left(-\phaz\right)=\st^{-1}\left(\phaz\right),\label{eq:scattering matrix behaviour-1}
\end{align}
where $\paulies 1$ is the first Pauli matrix, since $\pvity\left(\xl\right)$
satisfies Eq.\ \eqref{eq:modulation}. As a result, the product of
the modulus of the eigenvalues of $\sr\left(-\phaz\right)$ and $\sr\left(\phaz\right)$
must be $1$, and the same holds for $\st$. Therefore if $\sr\left(\phaz\right)$
has a pole then $\sr\left(-\phaz\right)$ has a zero, and hence $\st\left(-\phaz\right)$
has a zero too, which finally implies that $\st\left(\phaz\right)$
has a pole. In the next section, we analyze the dependency of the zeroes,
poles and EPs of $\sr$ and $\st$, and the scattering properties, on the
system parameters, and specifically the incident angle.

\section{Parametric study \label{sec:Examples}}

\begin{figure}[t]
\centering\sidesubfloat[]{\label{Fig:SCforPT}\includegraphics[scale=0.33]{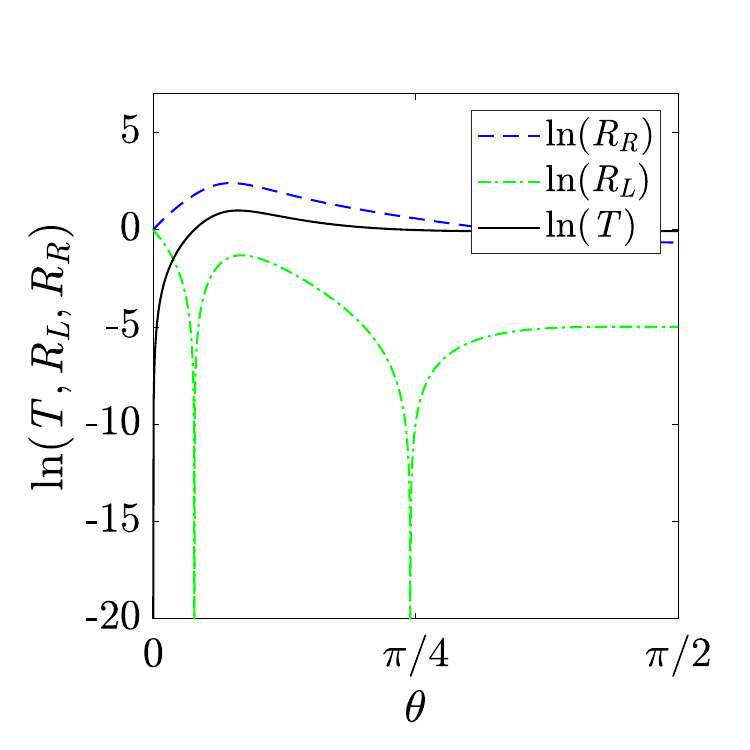}} \quad \centering\sidesubfloat[]{\label{Fig:StEVsforPT}\includegraphics[scale=0.33]{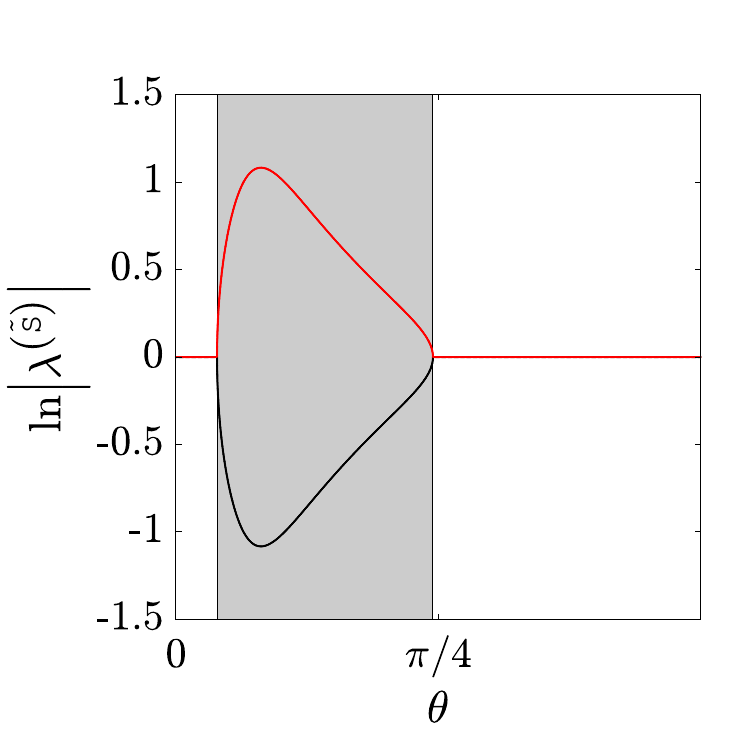}} \quad \centering\sidesubfloat[]{\label{fig:SrEVsforPT}\includegraphics[scale=0.33]{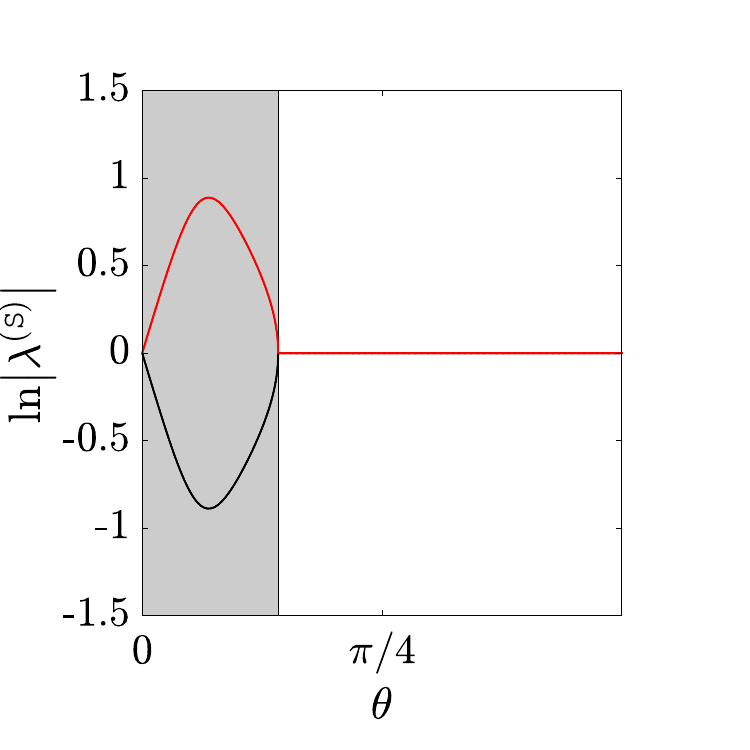}}

\caption {The case $\protect\phaz=0$. $\left(\text{a}\right)$ Logarithm of the transmittance $\protect\Tran=\left|\protect\trans\right|^{2}$
(solid black), and the two reflectances $\protect\RLorR=\left|\protect\rLorR\right|^{2}$
(dash-dotted green and dashed blue) as functions of the incident angle $\protect\ang$.
$\left(\text{b}\right)$ Logarithm of the magnitude of $\protect\evsc_{1,2}$,
as function of the incident angle $\protect\ang$. $\left(\text{c}\right)$
Logarithm of the magnitude of $\protect\evs_{1,2}$ as function of
the incident angle. In all panels, $\protect\eztep=10^{-1/2}$ and $ \protect\grating/\protect\ksh=2^{-1/2}$.}

{\small{}{}\label{fig:SCAndEVsOfSrSt}}{\small\par}
\end{figure}

In our case study, we set $\text{\ensuremath{\pvityh}}=4$, and study
the scattering properties as functions of the remaining parameters
of the system. We begin with Fig.\ \ref{Fig:SCforPT}, where we evaluate
the logarithm of the transmittance $\Tran$ (solid black), and the logarithm
of the two reflectances $\RLorR$ (dash-dotted green and dashed blue) as functions of
the incident angle $\ang$, setting the rest of the parameters to
\begin{equation}
\eztep\coloneqq\frac{\pvityp}{\pvityh}=10^{-1/2},\frac{\grating}{\ksh}=\frac{1}{\sqrt{2}},\phaz=0;\label{eq:parameters}
\end{equation}
and recall that since $\phaz=0$, the scatter is $\pt$-symmetric.
We observe that $\Tran$ varies from $0$ at $\ang=0$, to an anomalous
peak of $\Tran=2.63$ at $\ang=0.25$. Notably, the reflectance
from the right and left are different, where $\RL$ vanishes at $\ang=0.12$
and $\ang=0.77$, there the transmittance is unity. These angles correspond
to unidirectional reflection that occurs at the EPs of $\st$, as
was first reported by \citet{lin2011prl} for the case of normal incident
wave. To show this, we plot in Fig.\ \ref{Fig:StEVsforPT} the logarithm
of the magnitude of $\evsc_{1,2}$ as function of the incident angle
$\ang$. Indeed, we observe that there is a transition from unimodular
eigenvalues to non-unimodular eigenvalues at $\ang=0.12$ and $\ang=0.77$.

\citet{ge2012conservation} made the observation that while unidirectional
reflectivity occurs at the EPs of $\st$, the EPs of $\sr$ are those
that capture the breaking of the $\pt$ symmetry of the system. These
EPs are analyzed in Fig.\ \ref{fig:SrEVsforPT}, where we evaluate
the logarithm of the magnitude of $\evs_{1,2}$ as function of the
incident angle. We observe that for $0<\ang<0.44$, the eigenvalues
are non-unimodular, and are associated with broken $\pt$ symmetry. $\pt$ symmetry is restored
at $\ang=0.44$ , as the eigenvalues become
unimodular again. Specifically, both the eigenvalues
and eigenvectors coalesce at $\ang=0.44$, which identifies this angle as the EP of
$\sr$. Note that while the modulus of both the eigenvalues
is unity beyond this point, the eigenvalues themselves are different.

\floatsetup[figure]{style=plain,subcapbesideposition=top}

\begin{figure}[t]
\centering\sidesubfloat[]{{\small{}\label{fig:EPsOfSrAgainstOmega}}\includegraphics[scale=0.5]{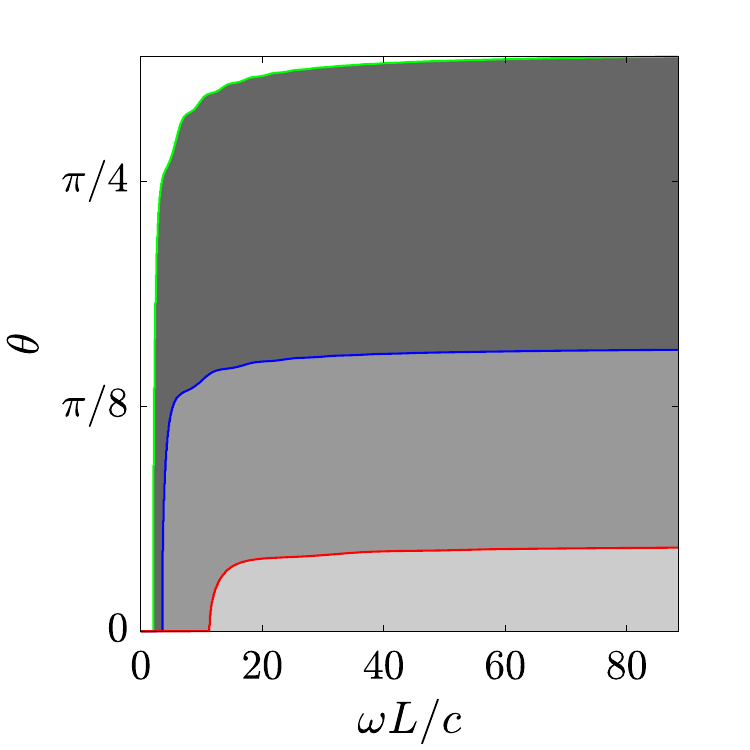}} \quad \centering\sidesubfloat[]{{\small{}\label{fig:EPsOfSrAgainstBeta}}\includegraphics[scale=0.5]{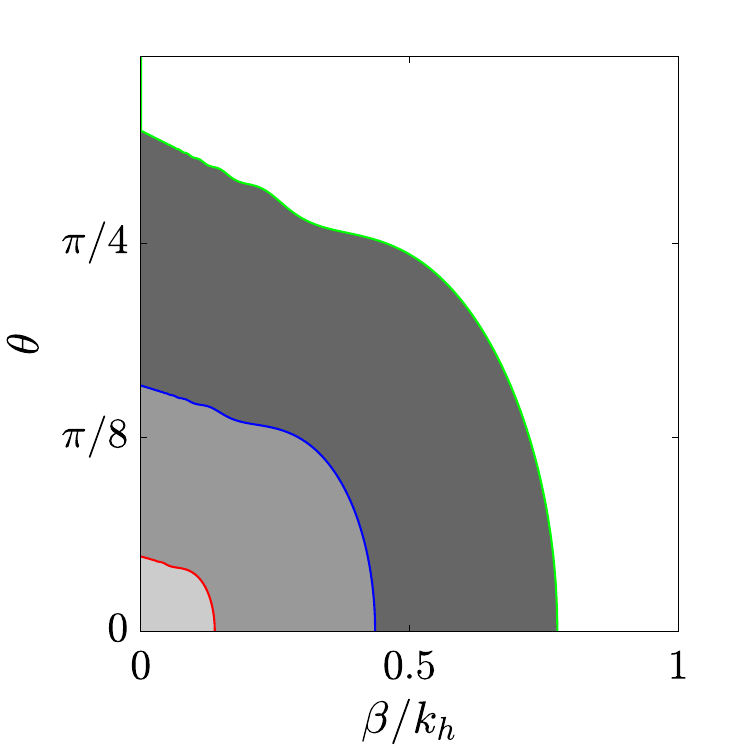}}

\caption{The case $\protect\phaz=0$. EPs of $\protect\sr$ for $\protect\eztep=10^{-2}$ (red), $10^{-1}$
(blue) and $10^{-1/2}$ (green). as functions of $\protect\ang$ versus
(a) $\protect\freq\protect\xr/\protect\light$, and (b) $\protect\grating/\protect\ksh$.
Light gray, gray and dark gray denote the broken phase region of $\protect\eztep=10^{-2}$,
$10^{-1}$ and $10^{-1/2}$, respectively.}

{\small{}{}\label{fig:EPsSr}}{\small\par}
\end{figure}

To examine the dependency of the $\pt$ symmetry breaking on the amplitude
of the modulation, we evaluate in Fig.\ \ref{fig:EPsSr} the EPs
of $\sr$ for $\eztep=10^{-2}$ (red), $10^{-1}$ (blue) and $10^{-1/2}$
(green), as functions of $\freq\xr/\light$ {[}panel (a){]}, and $\grating/\ksh$
{[}panel (b){]}. We highlight the regions that are associated with
broken $\pt$ symmetry when $\eztep=10^{-2}$, $10^{-1}$ and $10^{-1/2}$
by the light gray, gray and dark gray, respectively. We observe that
as we increase the amplitude of the perturbation, the region of the
broken phase is extended to lower frequencies, or equivalently greater
$\grating/\ksh$, and higher incident angles.

\floatsetup[figure]{style=plain,subcapbesideposition=top}

\floatsetup[figure]{style=plain,subcapbesideposition=top}

\begin{figure}[t]
\centering\sidesubfloat[]{\label{Fig:EPsOfStForPowerOfMinusTwo}\includegraphics[scale=0.33]{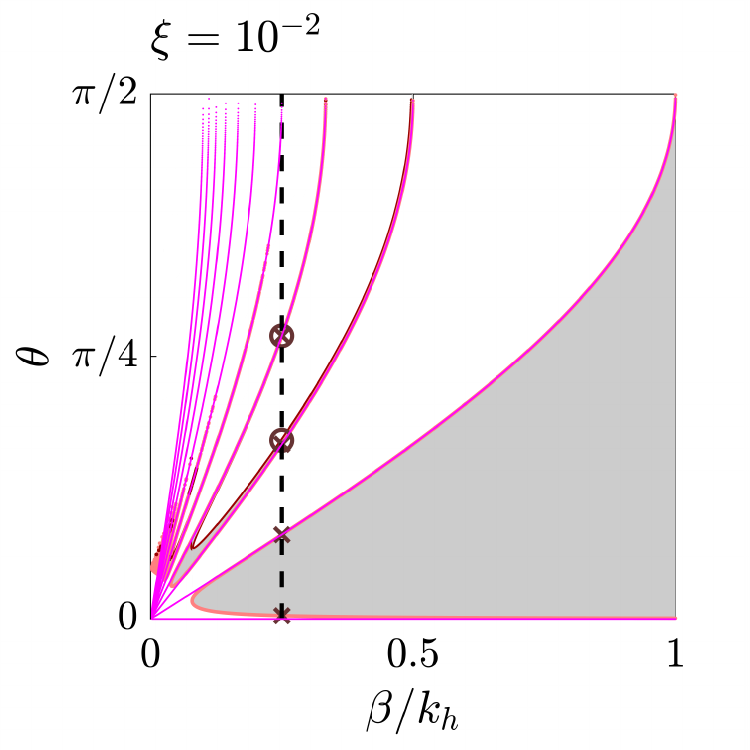}} \quad \centering\sidesubfloat[]{\label{Fig:EPsOfStForPowerOfMinusOne}\includegraphics[scale=0.33]{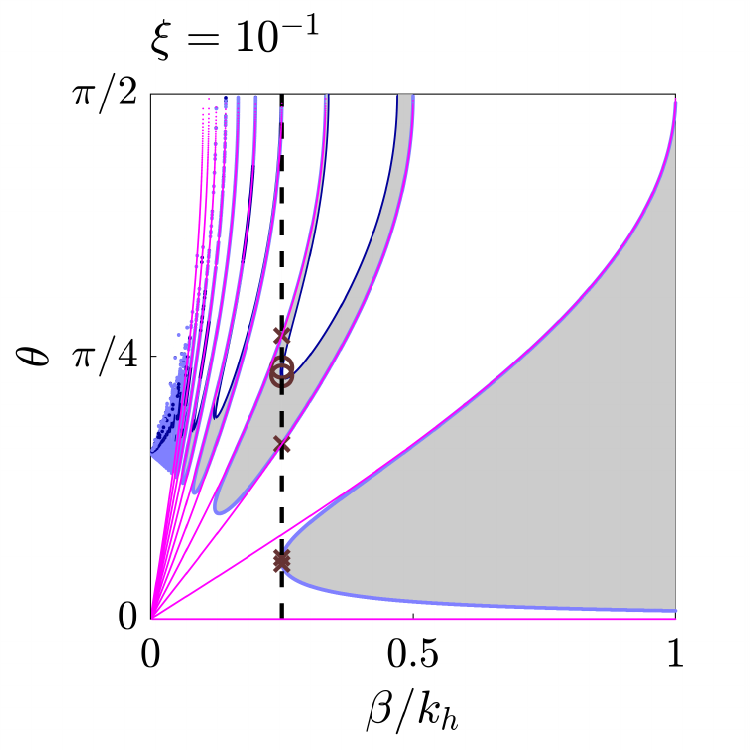}} \quad \centering\sidesubfloat[]{\label{fig:EPsOfStForPowerOfMinusHalf}\includegraphics[scale=0.33]{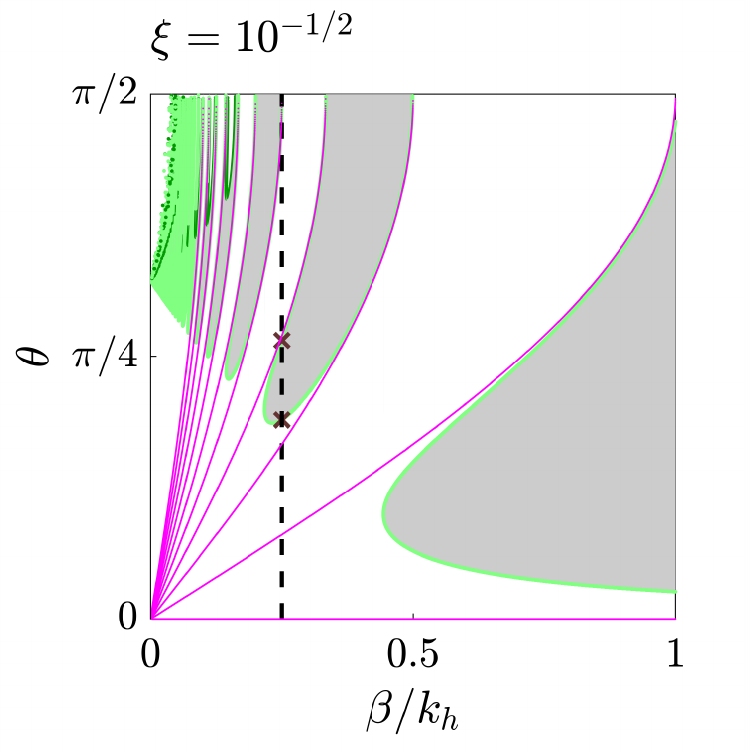}} \quad \centering\sidesubfloat[]{\label{fig:SCforpowerminustwo}\includegraphics[scale=0.33]{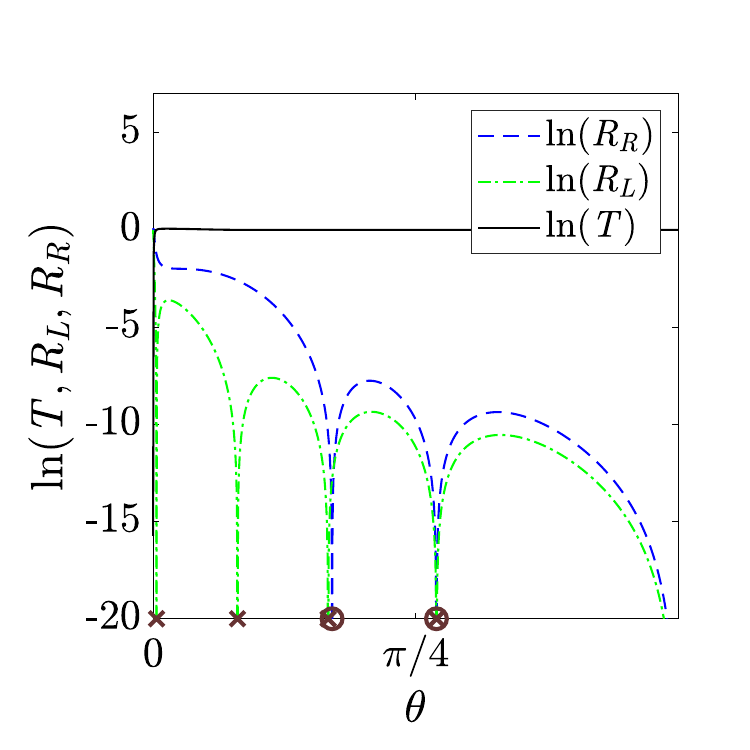}} \quad \centering\sidesubfloat[]{\label{fig:SCBDVForPowerOfMinusOne}\includegraphics[scale=0.33]{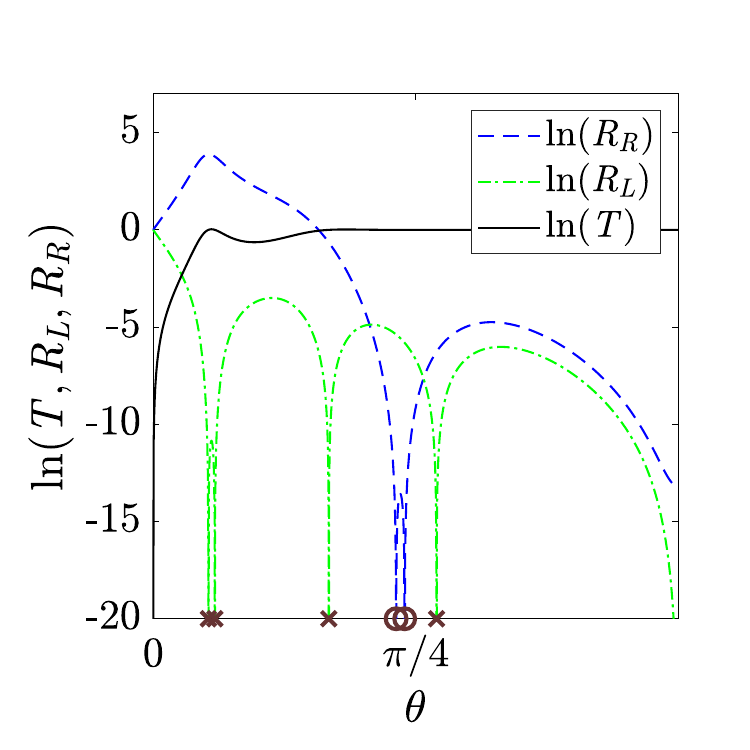}} \quad \centering\sidesubfloat[]{\label{fig:SCforpowerminushalf}\includegraphics[scale=0.33]{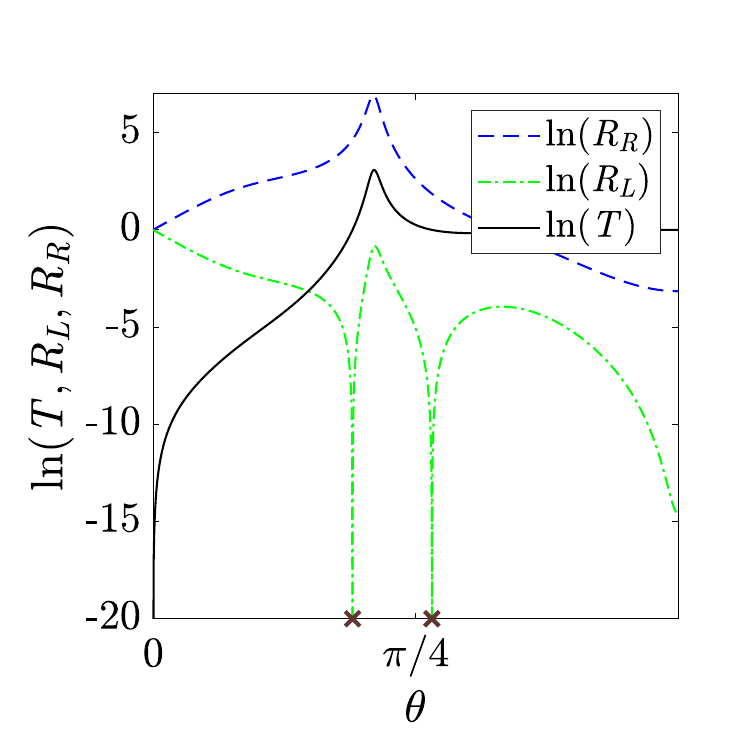}}

\caption{The case $\protect\phaz=0$. Phase diagram of $\protect\st$ in the space of $\left(\protect\ang,\protect\grating/\protect\ksh\right)$,
for $\left(\text{a}\right)$ $\protect\eztep=10^{-1/2}$, $\left(\text{b}\right)$
$\protect\eztep=10^{-1}$,$\left(\text{c}\right)$ $\protect\eztep=10^{-2}$.
The broken region is denoted in gray, and the EPs of $\protect\st$
are highlighted by the colored lines. Panels (d-f) show the logarithm
of $\Tran$ (solid black), and the two reflectances $\RLorR$ (dash-dotted green and
dashed blue) as functions of the incident angle $\ang$, for the three different
modulations, when setting $\grating/\ksh=1/4$.
}

{\small{}{}\label{fig:phasediagmst} }{\small\par}
\end{figure}
Having evaluated the phase diagram of $\sr$, we proceed to the phase
diagram of $\st$, shown in Fig.\ \ref{fig:phasediagmst}. Specifically,
panels $\left(\text{a}\right)$, $\left(\text{b}\right)$ and $\left(\text{c}\right)$
correspond to $\eztep=10^{-2}$ (red), $10^{-1}$ (blue) and $10^{-1/2}$
(green), respectively. Here, we distinguish between EPs associated
with $\refl$ and $\refr$ by light and dark shades, respectively.
In contrast with the phase diagram of $\sr$, which shows that there
is at most one EP for a given $\eztep$ and $\grating/\ksh$, here
there could be multiple EPs. For example, there are six EPs when $\eztep=10^{-1}$ and
$\grating/\ksh=0.32$ (black vertical line in Fig.\ \ref{Fig:EPsOfStForPowerOfMinusOne}).

In all the panels, there is a broken region that extends to $\grating/\ksh=1$.
Interestingly, this region is bounded from above by the function $\grating=\ksh\sin\ang$,
which corresponds to the first Fabry-P\'erot
resonance \cite{hodgson2005optical}. In
the \hyperref[Appendix-A]{Appendix}, we have used a multiple scale expansion to derive an approximated solution, under the assumption $\grating\approx\ksh\sin\ang$ and that the
perturbation is small,
resembling the case studied by \citet{lin2011prl}; using this approximation,
we show that indeed the scatterer exhibits unidirectional reflection
under these assumptions. In fact, all Fabry-P\'erot
resonances, i.e., $\nnb\grating=\ksh\sin\ang$ for any $m\in\mathbb{N}$,
provide an approximation for the EPs of $\st$. These approximations
for $\nnb=0..10$ are depicted in magenta lines.

The quality of the Fabry-P\'erot resonances approximation depends on how small
the perturbation and detuning are. To show this, we use our analytical
expressions for $\refl$ and $\refr$ {[}Eqs.\ \eqref{eq:transfer mat}-\eqref{eq:sdef coefficients}{]}
to find that they are proportional to 
\begin{align}
\refl\propto\besseli_{-\left(\order+1\right)}\left(\order\sqrt{\epsa}\right) & ,\quad\refr\propto\besseli_{-\left(\order-1\right)}\left(\order\sqrt{\epsa}\right),\label{eq:vanish-1}
\end{align}
where $\besseli$ is the modified Bessel function of the first kind,
and we recall that $\order=\left(\ksh/\grating\right)\sin\ang$. The
modified Bessel function vanishes in the limit $\epsa\rightarrow0^{+}$
when its order is a nonzero integer, i.e., $\order\rightarrow\nnb\neq1$,
in line with our assumptions. When $\order$ is close to an even (odd)
number from above (below), then $\besseli_{-\left(\order-1\right)}$
and $\besseli_{-\left(\order+1\right)}$ vanish for some small positive
number. This corresponds to vanishing $\refl$ and $\refr$ at some
small perturbation amplitude $\epsa$, near a Fabry-P\'erot resonance of even
(odd) integer. Exceptions for this rule are for (\emph{i})
$\besseli_{0}$, which is nonzero at $\epsa=0$, hence $\refr$ does
not vanish in the limit $\order\rightarrow1$ that is associated
with the first Fabry-P\'erot resonance $\left(m=1\right)$; (\emph{ii})
when $\order\rightarrow0^{+}$, $\refr$ is proportional to $\besseli_{1^{-}}$,
which does not vanish at some small positive number, hence $\refr$
does not vanish for $m=0$.

Indeed, we observe that for smallest modulation (Fig.\ \ref{Fig:EPsOfStForPowerOfMinusTwo}),
almost all of the magenta lines coincide with the EPs that we find
using the exact solution, including the zeroth order, i.e., when $\ang=0$.
By contrast, for the largest modulation (Fig.\ \ref{fig:EPsOfStForPowerOfMinusHalf}),
the zeroth order does not provide a good approximation at all, and
the first order begins to coincide with the exact solution only from
$\grating/\ksh\approx0.5$. 

By analyzing how the zeros of the modified Bessel functions depend
on the order, we deduce that the EPs associated with vanishing $\refl$
will always be closer to the Fabry-P\'erot resonances than those of vanishing
$\refr$. Since higher Fabry-P\'erot resonances become closer in the
$\left(\grating/\ksh,\ang\right)$ space, and the approximated solution
improves at smaller modulation amplitude and higher order, the EPs
of $\refl$ and $\refr$ become closer at smaller modulations and
higher   Fabry-P\'erot resonances; from a certain frequency, they practically
coincide. Thus, this trend provides guidelines for
achieving bidirectional zero-reflection. 

\floatsetup[figure]{style=plain,subcapbesideposition=top}

\begin{figure}[t]
\centering\sidesubfloat[]{\label{fig:SCforPhaseOfOneToTwo-1}\includegraphics[scale=0.33]{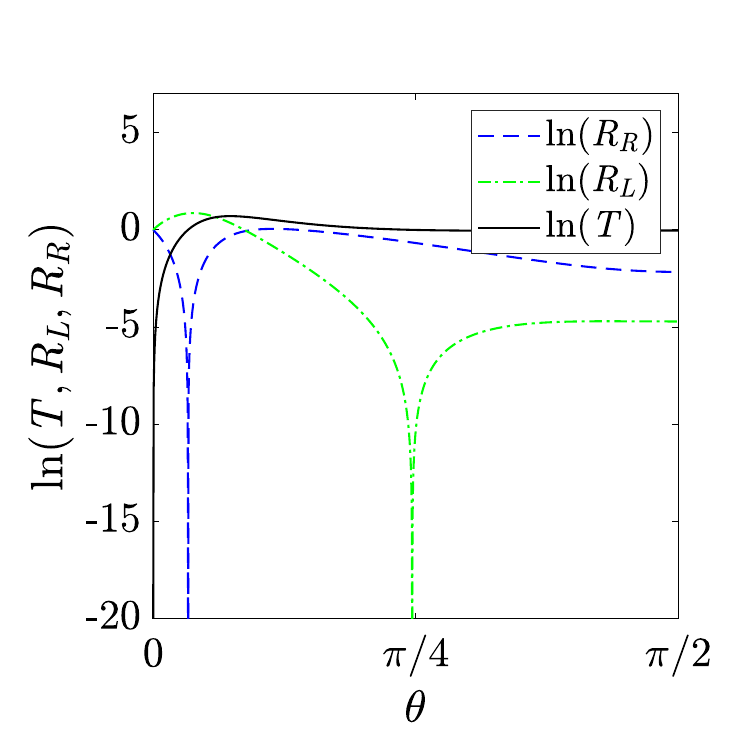}} \quad \centering\centering\sidesubfloat[]{\label{Fig:StPTPhaseForPhaseOfOneToTwo}\includegraphics[scale=0.33]{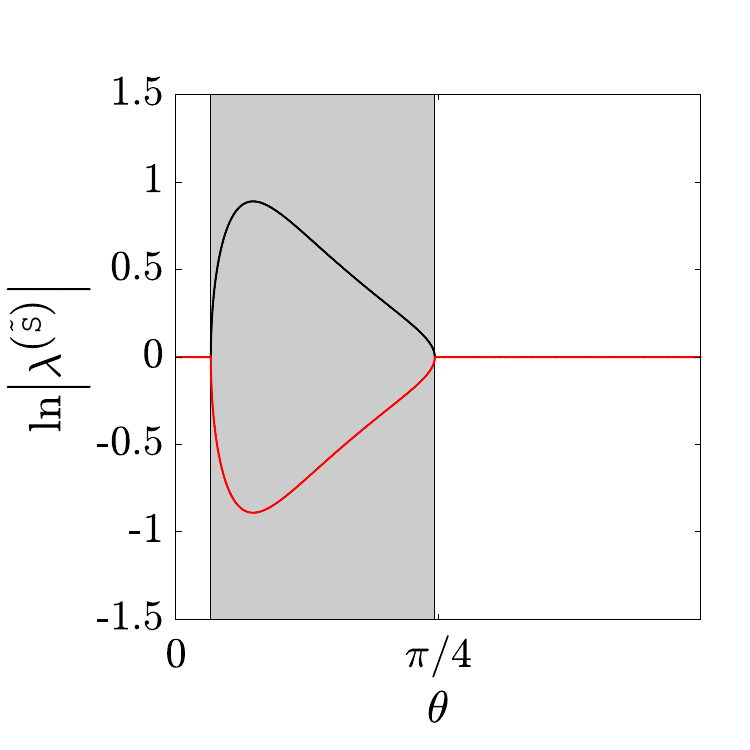}} \quad \centering\sidesubfloat[]{\label{fig:SrPTPhaseForPhaseOfOneToTwo}\includegraphics[scale=0.33]{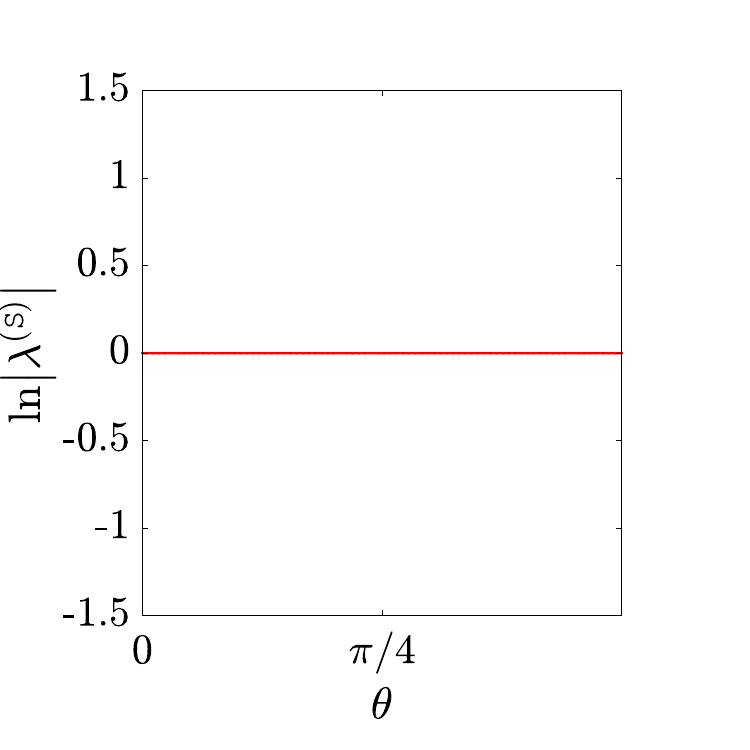}}

\caption{The case $\protect\phaz=\xr/2$. $\left(\text{a}\right)$ Logarithm of the transmittance $\protect\Tran=\left|\protect\trans\right|^{2}$
(solid black), and the two reflectances $\protect\RLorR=\left|\protect\rLorR\right|^{2}$
(dash-dotted green and dashed blue) as functions of the incident angle $\protect\ang$.
$\left(\text{b}\right)$ Logarithm of the magnitude of $\protect\evsc_{1,2}$,
as function of the incident angle $\protect\ang$. $\left(\text{c}\right)$
Logarithm of the magnitude of $\protect\evs_{1,2}$ as function of
the incident angle. In all panels, $\protect\eztep=10^{-1/2}$ and $ \protect\grating/\protect\ksh=2^{-1/2}$.}

{\small{}{}\label{fig:PTPhaseWP}}{\small\par}
\end{figure}

To demonstrate this, we evaluate in the remaining panels the logarithm
of $\Tran$ (solid black), and the two reflectances $\RLorR$ (dash-dotted green and
dashed blue) as functions of the incident angle $\ang$, for the three different
modulations, when setting $\grating/\ksh=1/4$. Specifically, panels
$\left(\text{d}\right)$, $\left(\text{e}\right)$ and $\left(\text{f}\right)$
correspond to $\eztep=10^{-2}$, $10^{-1}$ and $10^{-1/2}$, respectively.
Panel (d) shows that $\RL$ vanishes at $\ang=0.01,0.25,0.52$ and
$0.85$, as denoted by the cross marks, while $\RR$ vanishes very
closely to the latter two angles, i.e., near $\ang=0.52$ and $0.85$,
as designated by the circle marks. This observation is in agreement
with panel (a), where the wavenumber ratio $\grating/\ksh=1/4$ is
denoted by the dashed line. This line intersects the exceptional line
that is near the zeroth Fabry-P\'erot resonance at $\ang=0.01$, and then
intersects the exceptional lines that practically coincide with the
 Fabry-P\'erot resonances that are defined by $m=$1, 2 and 3. Indeed, near Fabry-P\'erot resonances for which $m$ equals zero and one, there is no EP that is associated with $\RR$,
while near the higher order Fabry-P\'erot resonances, i.e., two and three, there are
pairs of EPs of $\RL$ and $\RR$ that are very close to each other
about $\ang=0.52$ and $0.85$. Panel (e) shows that $\RL$ vanishes
at $\ang=0.17,0.18,0.53$ and $0.85$; here, since the modulation
amplitude is greater than in panel (d), the angles at which $\RR$
vanishes, namely, $\ang=0.73$ and $0.75$, are not as close to the
nearest angles at which $\RL$ vanishes. Again, these EP observations
agree with the way in which the dashed line that
denotes $\grating/\ksh=1/4$ intersects the exceptional lines and
the Fabry-P\'erot resonances. Finally, we see that in Panel (f), which corresponds
to the larger modulation amplitude, there are no angles at which $\RR$
vanishes, nor vanishing values $\RL$ near the two lowest-order Fabry-P\'erot resonances (zero and one), but only from order two.

\begin{figure}[t]
\centering\sidesubfloat[]{\label{fig:SrPTPhaseForPhaseOfTwoWPAgainstFreq}\includegraphics[scale=0.33]{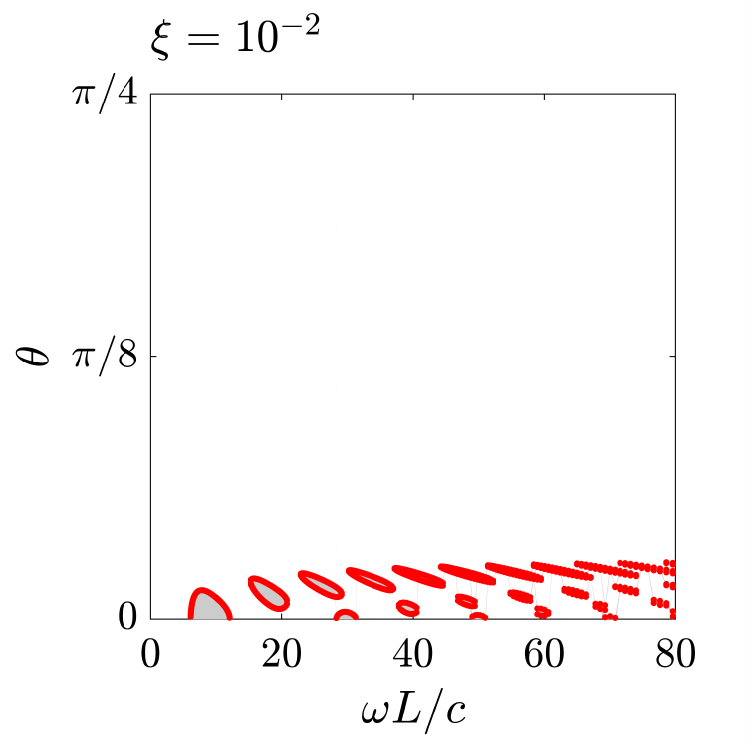}} \quad \centering\sidesubfloat[]{\label{fig:SrPTPhaseForPhaseOfOneWPAgainstFreq}\includegraphics[scale=0.33]{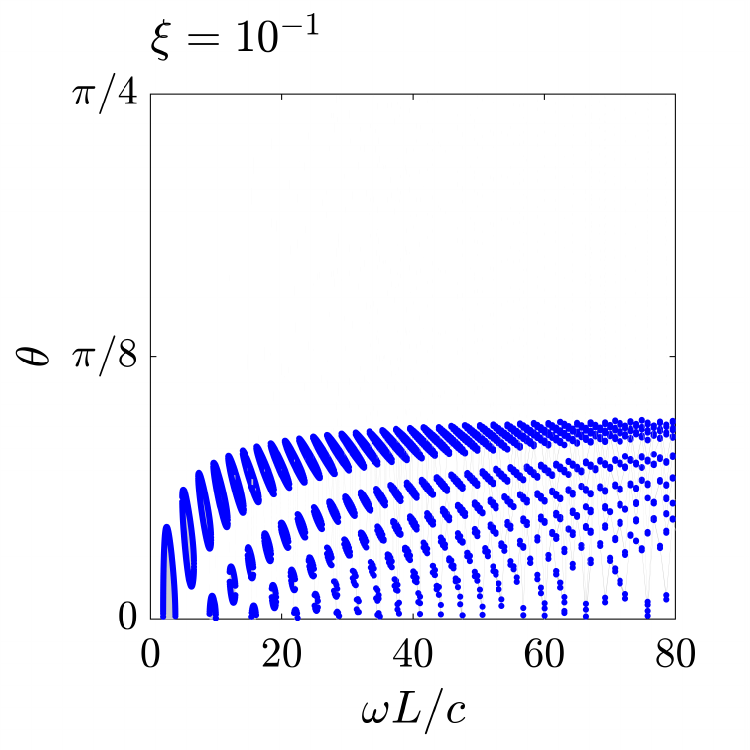}} \quad \centering\sidesubfloat[]{\label{fig:SrPTPhaseForPhaseOfHalfWPAgainstFreq}\includegraphics[scale=0.33]{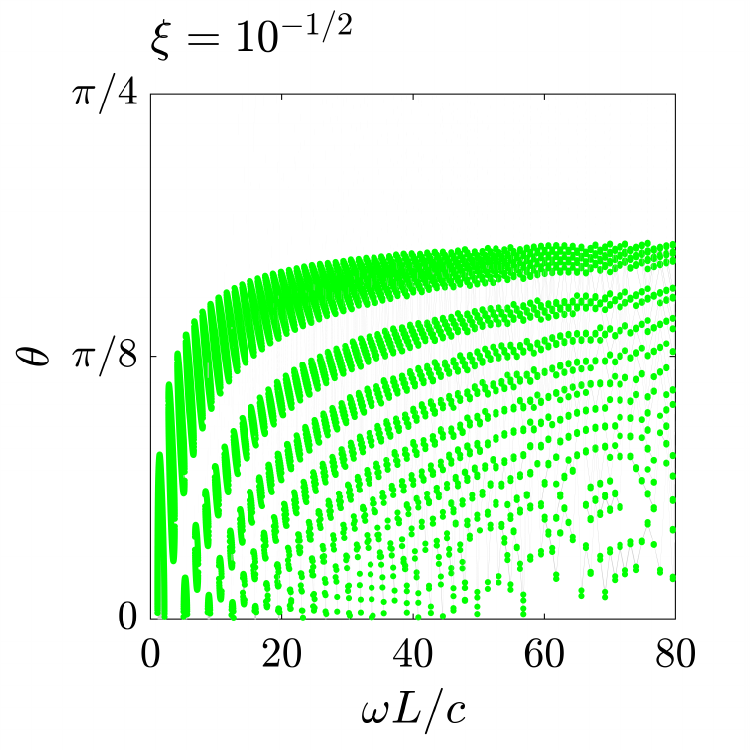}} \quad \centering\sidesubfloat[]{\label{fig:SrPTPhaseForPhaseOfOneToTwoWP}\includegraphics[scale=0.33]{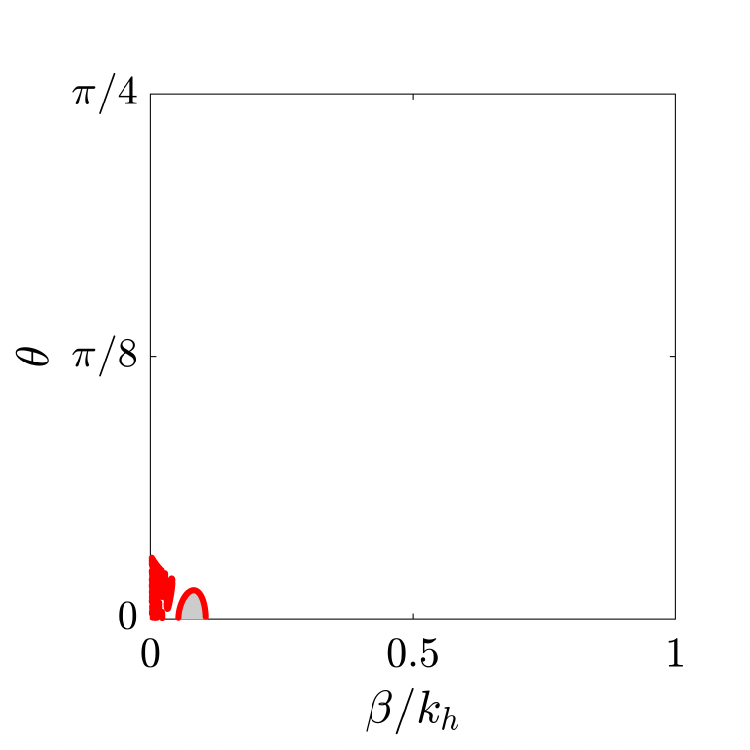}} \quad \centering\sidesubfloat[]{\label{fig:SrPTPhaseForPhaseOfOneToOneWP}\includegraphics[scale=0.33]{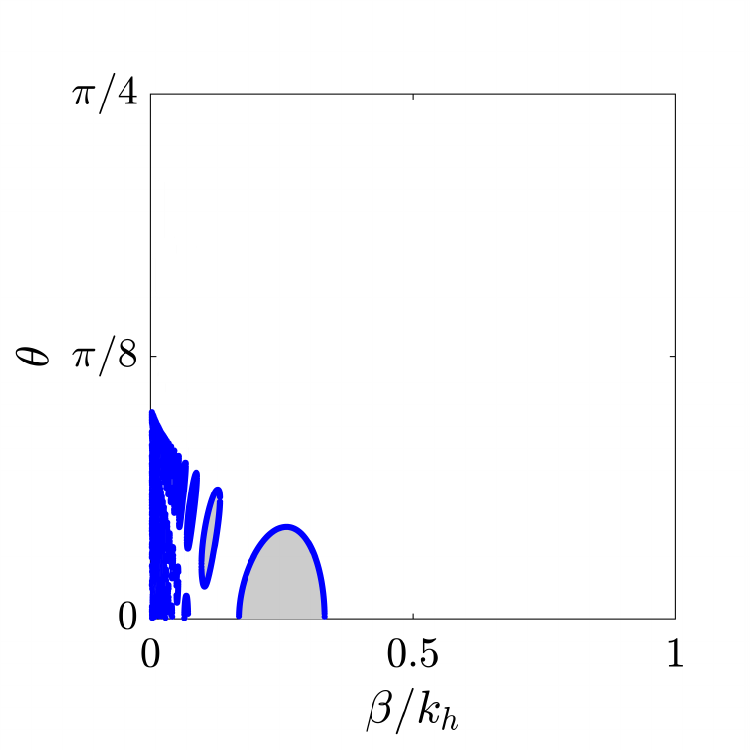}} \quad \centering\sidesubfloat[]{\label{fig:SrPTPhaseForPhaseOfOneToHalfWP}\includegraphics[scale=0.33]{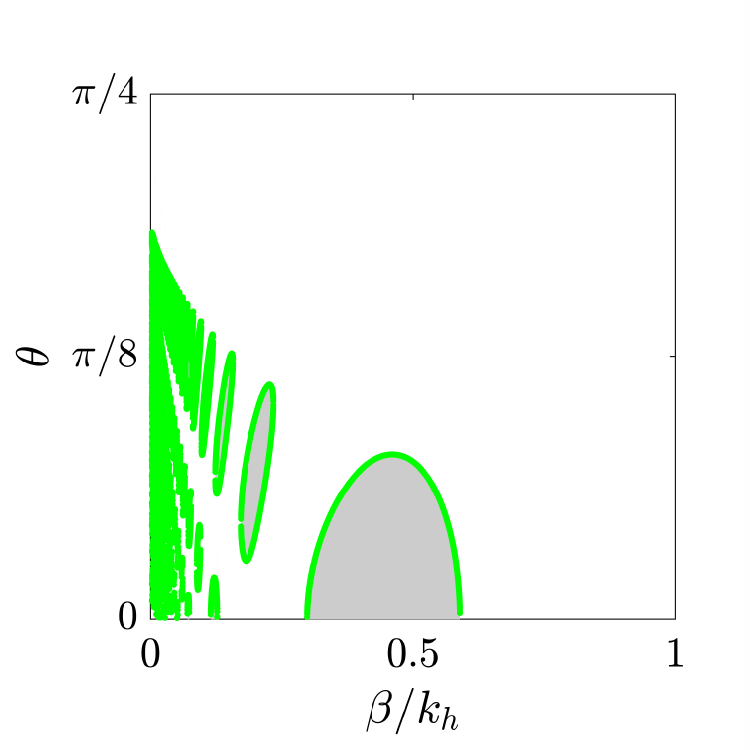}}\caption{The case $\phaz=\xr/2$. Phase diagram of $\protect\sr$ in the $\left(\ang,\freq\xr/\light\right)$ space {[}panels (a-c){]} and
$\left(\protect\ang,\protect\grating/\protect\ksh\right)$ space {[}panels (d-f){]} ,
for $\left(\text{a,d}\right)$ $\protect\eztep=10^{-1/2}$, $\left(\text{b,e}\right)$
$\protect\eztep=10^{-1}$ and $\left(\text{c,f}\right)$ $\protect\eztep=10^{-2}$.
The broken region is denoted in gray, and the EPs of $\protect\sr$
are highlighted by the colored lines.}

{\small{}{}\label{fig:ELOfSr}}{\small\par}
\end{figure}

The analysis so far was for a zero modulation phase, yielding a $\pt$-symmetric
scatterer. We recall that when $\phaz=\xr/2$, the scatterer is also
$\pt$-symmetric: this case is analyzed next.  We begin with Fig.\ \ref{fig:PTPhaseWP},
which is the same as Fig.\ \ref{fig:SCAndEVsOfSrSt}, only for $\phaz=\xr/2$.
 A comparison between panels \ref{Fig:SCforPT} and \ref{fig:SCforPhaseOfOneToTwo-1}
shows that while only $\RL$ vanishes (twice) when $\phaz=0$, when
$\phaz=\xr/2$ both $\RL$ and $\RR$ vanish once, at different angles.
A comparison of the remaining two panels in Figs.\ \ref{fig:SCAndEVsOfSrSt} and \ref{fig:PTPhaseWP}
shows that while the broken region of
$\st$ is similar for $\phaz=0$ and $\xr/2$, the broken region of
$\sr$ is completely different, since when $\phaz=\xr/2$, $\sr$
is always at the $\pt$-symmetric phase for the same given scattering
parameters. A more complete picture of the difference in the phase diagrams of the two $\pt$-symmetric systems is given next, by providing the diagrams that are associated with $\phaz=\xr/2$. We start with Fig.\ \ref{fig:ELOfSr}, which shows the 
EPs of $\sr$ as functions of $\ang$ versus 
$\freq\xr/\light$ {[}panels (a-c){]}, and  $\grating/\ksh$ {[}panels (d-f){]} for
$\eztep=10^{-2}$ {[}red, panels (a) and (d){]}, $10^{-1}$ {[}blue, panels (b) and (e){]} and $10^{-1/2}$  {[}green, panels (c) and (f){]}, 
where the broken regions are highlighted with gray.  A comparison with the diagram of  $\phaz=0$ (Fig.\ \ref{fig:EPsSr}) shows that $\phaz=\xr/2$ has a much richer diagram, exhibiting multiple re-entries to the broken region when either one of the three parameters ($\ang$, $\freq$ and $\grating$) is varied. This is in sharp contrast with a single entry when $\phaz=0$. In addition, when $\phaz=0$, the broken region of the greater modulations encloses the broken region of the smaller modulations, while when $\phaz=\xr/2$, there are regions that belong to the broken phase of  smaller modulations, while belonging to the exact phase of greater modulations.
\floatsetup[figure]{style=plain,subcapbesideposition=top}

\begin{figure}[t]
\centering\sidesubfloat[]{\label{fig:EPsOfStForPowerOfMinusTwoWP}\includegraphics[scale=0.33]{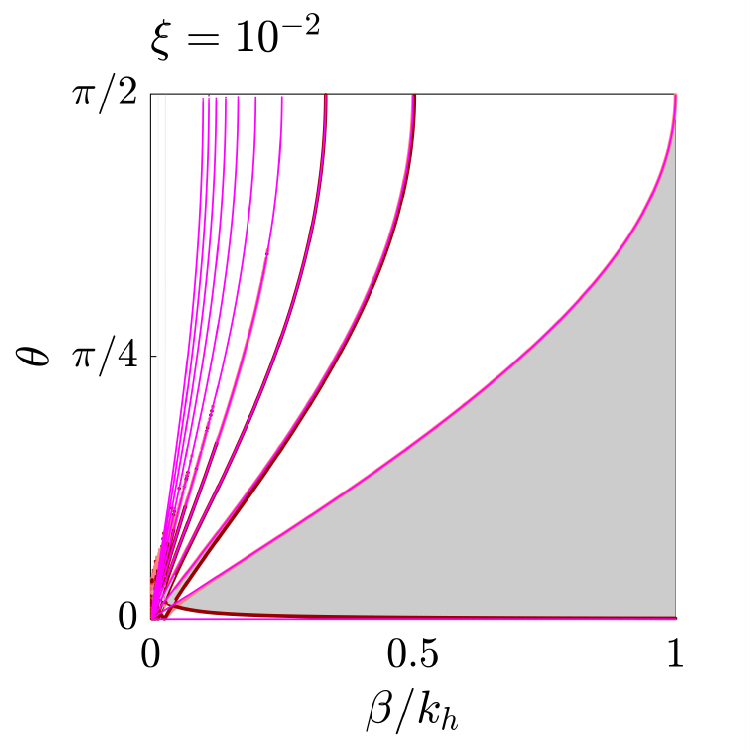}} \quad \centering\sidesubfloat[]{\label{fig:EPsOfStForPowerOfMinusOneWP}\includegraphics[scale=0.33]{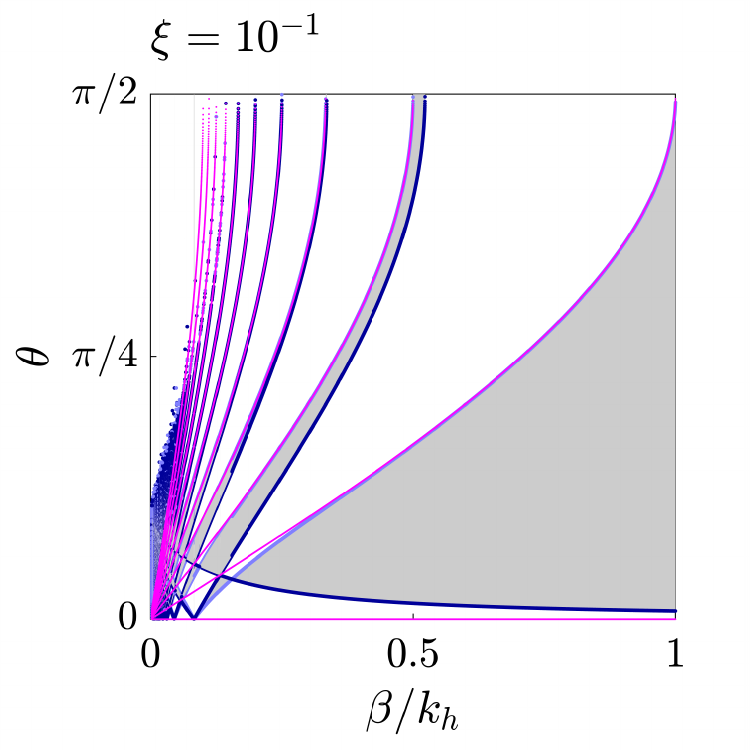}} \quad \centering\sidesubfloat[]{\label{fig:EPsOfStForPowerOfMinusHalfWP}\includegraphics[scale=0.33]{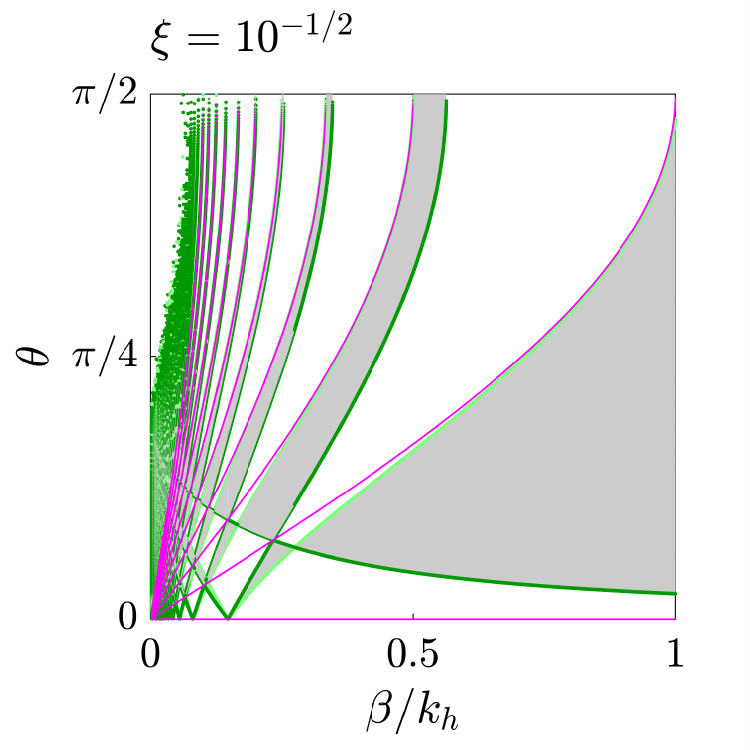}}

\caption{Phase diagram of, $\protect\st\left(\protect\phaz/\protect\xr=1/2\right)$
in the space of $\left(\protect\ang,\protect\grating/\protect\ksh\right)$,
for $\left(\text{a}\right)$ $\protect\eztep=10^{-1/2}$, $\left(\text{b}\right)$
$\protect\eztep=10^{-1}$,$\left(\text{c}\right)$ $\protect\eztep=10^{-2}$.
The broken region is denoted in gray, and the EPs of $\protect\st$
are highlighted by the colored lines.}

{\small{}{}\label{fig:ELOfSt}}{\small\par}
\end{figure}

In Fig.\ \ref{fig:ELOfSt} we also evaluate the phase diagram of $\st$ in the  ($\ang,\grating/\ksh$) space;   panels (a-c)
correspond to $\eztep=10^{-2}$, $10^{-1}$ and $10^{-1/2}$, where the EPs are highlighted in red, blue and green, respectively, and the broken regions are in gray.  We distinguish between the EPs that are associated
with zero $\refl$ and $\refr$ by light and dark shades, respectively. There is some similarity with the diagram of $\phaz=0$, however here the structure is more complicated and the broken regions are larger.

Here again, all Fabry-P\'erot resonances provide an approximation for the EPs of
$\st$, and these approximations for $\nnb=0..10$ are depicted in magenta
lines. Now we find that the
expressions for $\refl$ and $\refr$ are proportional to the Bessel functions \begin{align}
\refl\propto\bessel_{-\left(\order+1\right)}\left(\order\sqrt{\epsa}\right) & ,\quad\refr\propto\bessel_{-\left(\order-1\right)}\left(\order\sqrt{\epsa}\right)\bessel_{\left(\order-1\right)}\left(\order\sqrt{\epsa}\right).\label{eq:vanishWP}
\end{align}
In contrast with the behavior of the modified Bessel functions in Eq.\ \eqref{eq:vanish-1}, the Bessel functions in Eq.\ \eqref{eq:vanishWP} vanish at some positive argument when $\order$ approaches to a positive integer number from below. This corresponds to vanishing
$\refl$ and $\refr$ at some small perturbation amplitude $\epsa$,
near a Fabry-P\'erot resonance. Here there is only one exception to this
rule, namely, that $\refr$ does not vanish
in the limit $\order\rightarrow1$, which is associated with the Fabry-P\'erot resonance, i.e., $m=1$.

\floatsetup[figure]{style=plain,subcapbesideposition=top}

\floatsetup[figure]{style=plain,subcapbesideposition=top}

\begin{figure}[t]
\centering\sidesubfloat[]{\label{fig:SCforPhaseOfOneToTwo}\includegraphics[scale=0.33]{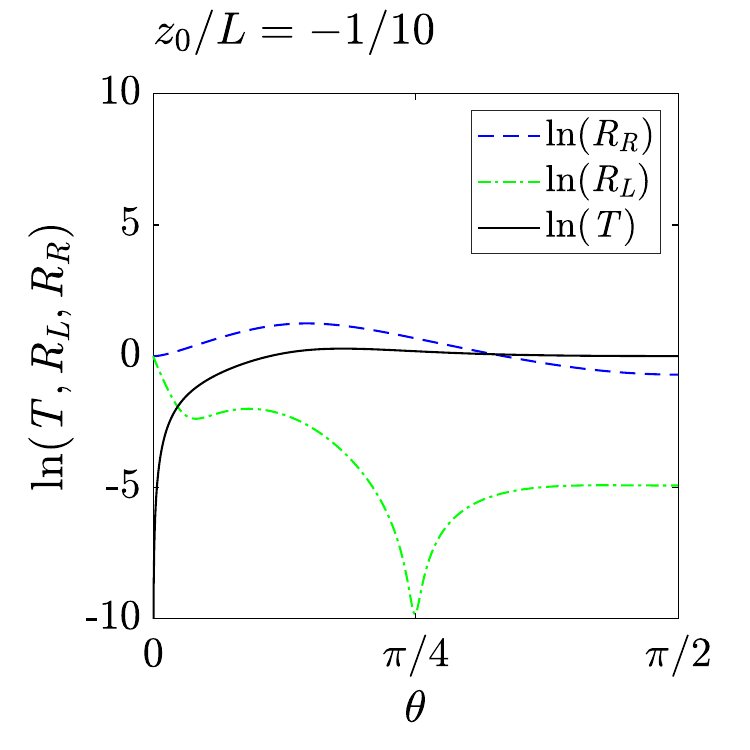}} \quad \centering\sidesubfloat[]{\label{Fig: SCforPhaseOfThreeToTen}\includegraphics[scale=0.33]{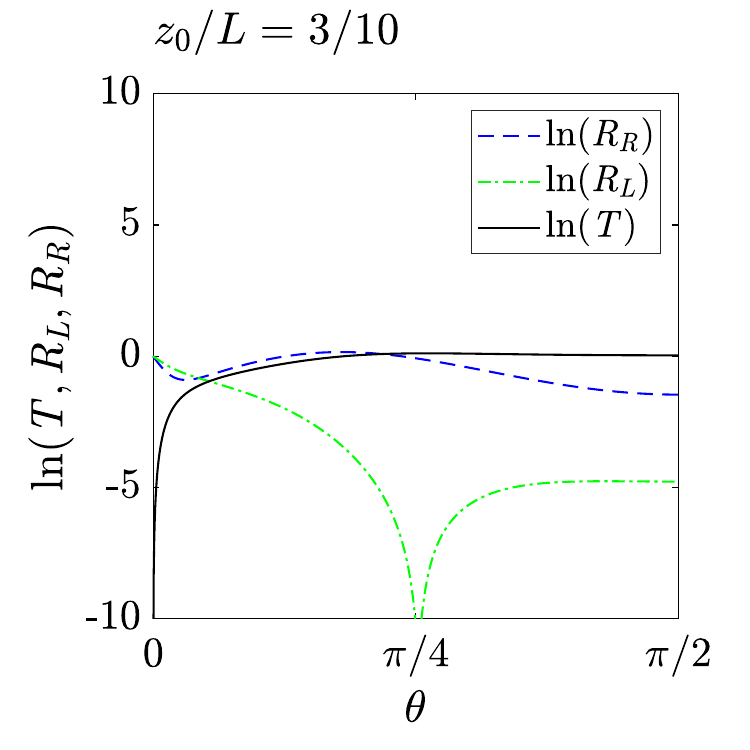}} \quad \centering\sidesubfloat[]{\label{Fig:SCforPhaseOfMinusTwoToFive}\includegraphics[scale=0.33]{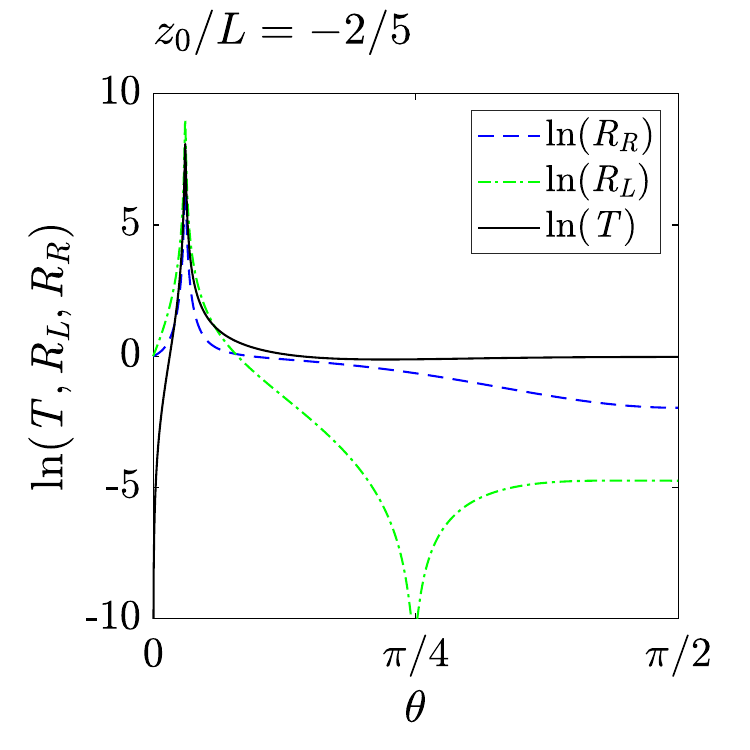}}

\caption{The transmittance $\protect\Tran=\left|\protect\trans\right|^{2}$
(solid black), reflectance from the left $\protect\RL=\left|\protect\refl\right|^{2}$
(dash-dotted green) and reflectance from the right $\protect\RR=\left|\protect\refr\right|^{2}$
(dashed blue) as functions of the incident angle, for $\protect\eztep=10^{-1/2}$,
$\protect\grating/\protect\ksh=2^{-1/2}$, and $\protect\phaz/\protect\xr$
equals (a) $-1/10$, (b) $3/10$ and (c) $-2/5$. }

{\small{}{}\label{fig:TRRWP}}{\small\par}
\end{figure}

We proceed to analyze scatterers that do not satisfy $\pt$ symmetry, by considering three exemplary values of $\phaz$ that are different from 0 and $\xr/2$.  This is carried out in Fig.\ \ref{fig:TRRWP}, where we evaluate the logarithm of $\Tran$ and
$\RLorR$ as functions of $\ang$,
setting $\eztep=10^{-1/2}$ and $\grating/\ksh=2^{-1/2}$. Specifically,
panels $\left(\text{a}\right)$, $\left(\text{b}\right)$ and $\left(\text{c}\right)$
correspond to $\phaz=-1/10,3/10$ and $-2/5$, respectively. While $\RL$ vanish near $\ang=\pi/4$ in all the panels (corresponding to $\order=1$), 
overall there is a strong variation in the scattering properties from one
$\phaz$ to another. For example, when $\phaz/\xr=-1/10$, $\Tran$
peaks at $\ang=0.57$ to $1.34$, and then decays to $1.01$ at $\ang=\pi/2$.
By contrast, when $\phaz/\xr=3/10$, $\Tran$ peaks at $\ang=0.82$
to $1.11$, and then decays to $1.04$ at $\ang=\pi/2$. Furthermore,
when $\phaz/\xr=-1/10$, $\RL$ is unity at $\ang=0$, decays to
$0.09$ at $\ang=0.13$ which is a local minimum, then peaks to $0.14$
at $\ang=0.29$, and then decays to zero at $\ang=0.78$, and increases
again to $0.01$ at $\ang=\pi/2$. By contrast, when $\phaz/\xr=3/10$,
the peaks of $\RL$ is at $\ang=0$, from which it decays to zero
at $\ang=0.79$ and increases again to $0.01$ at $\ang=\pi/2$. The
most interesting result is observed for $\phaz/\xr=-2/5$ about $\ang=0.09$,
where $\Tran,\RR$ and $\RL$ all peak to giant values at the order
of $10^{4}.$ 

\begin{figure}[t]
\centering\sidesubfloat[]{\label{Fig:PolesForPowerOfMinusTw}\includegraphics[scale=0.33]{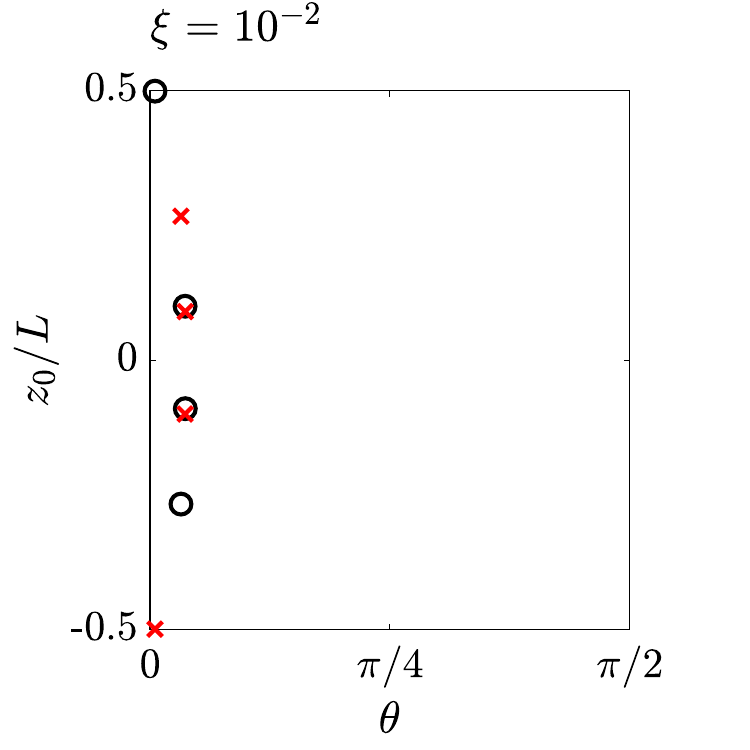}} \quad \centering\sidesubfloat[]{\label{Fig:PolesForPowerOfMinusOne}\includegraphics[scale=0.33]{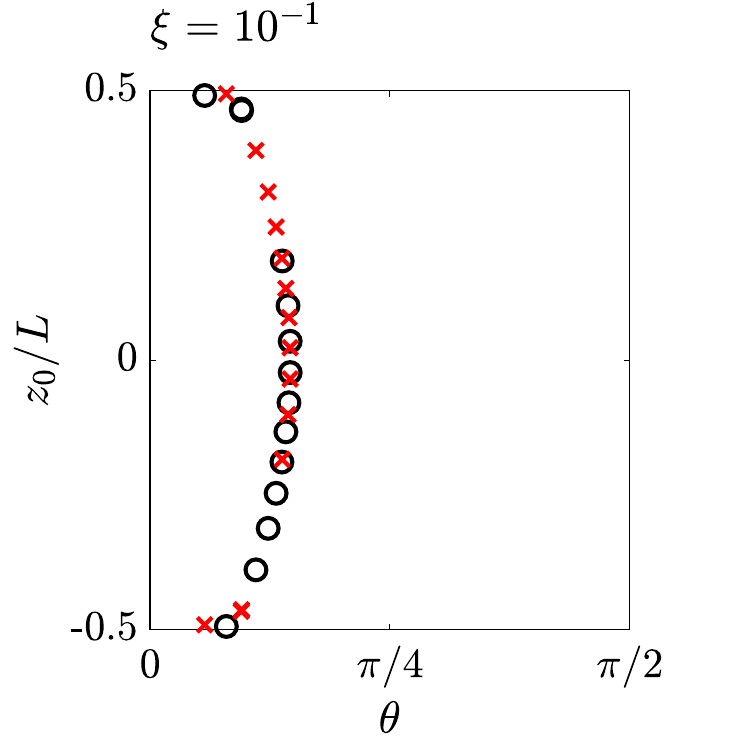}} \quad \centering\sidesubfloat[]{\label{fig:PolesForPowerOfMinusHalf}\includegraphics[scale=0.33]{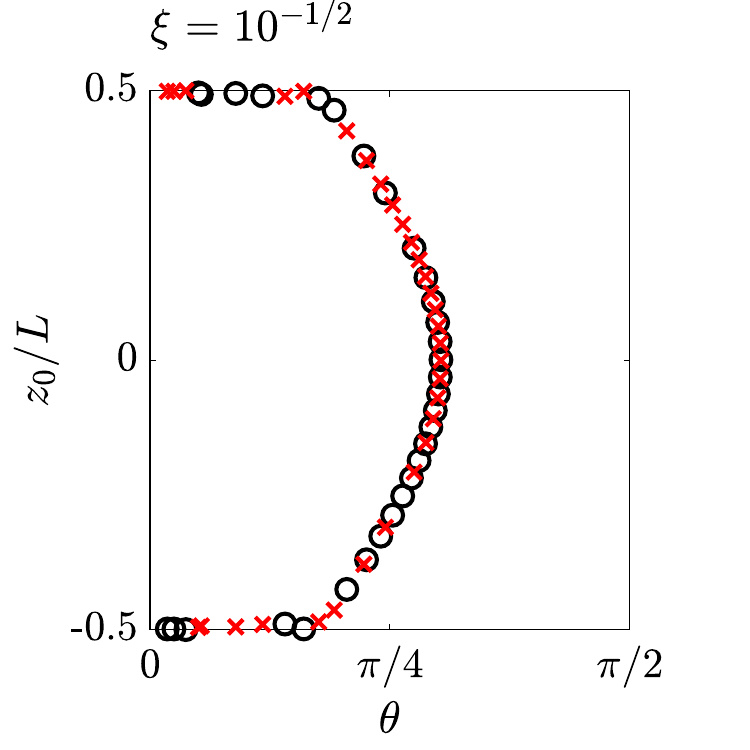}} \quad \centering\sidesubfloat[]{\label{Fig:SrPTPhaseForPowerOfMinusTwo}\includegraphics[scale=0.33]{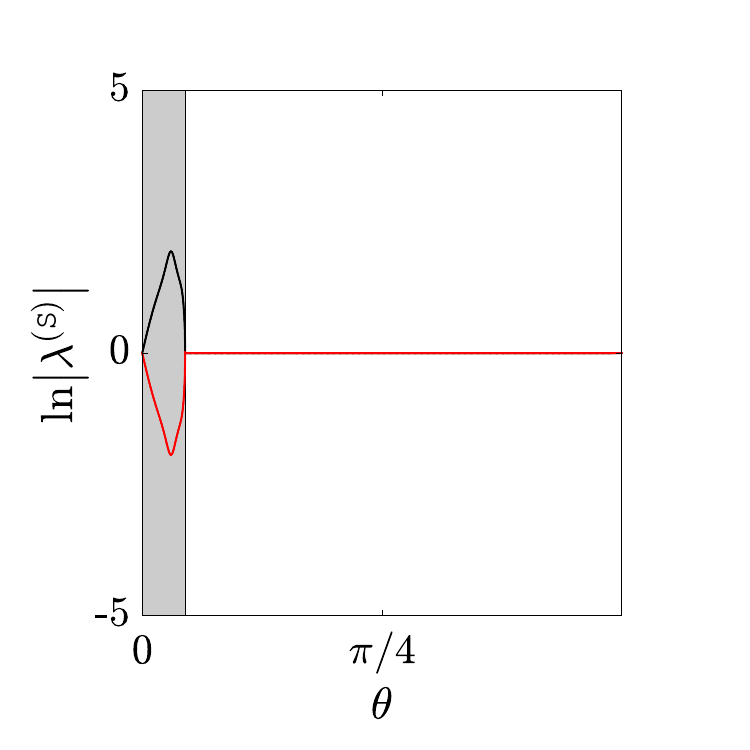}} \quad \centering\sidesubfloat[]{\label{Fig:SrPTPhaseForPowerOfMinusOne}\includegraphics[scale=0.33]{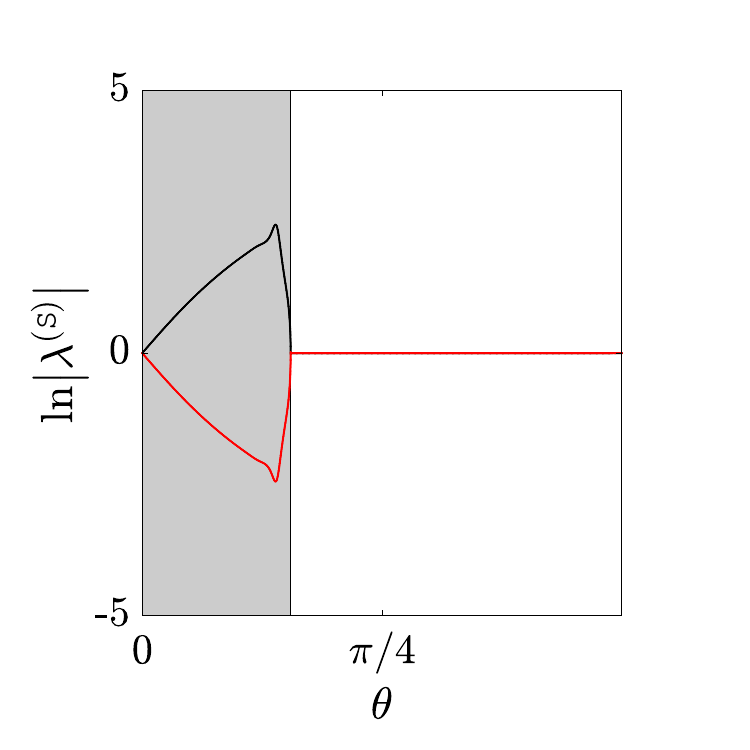}} \quad \centering\sidesubfloat[]{\label{Fig:SrPTPhaseForPowerOfMinusHalf}\includegraphics[scale=0.33]{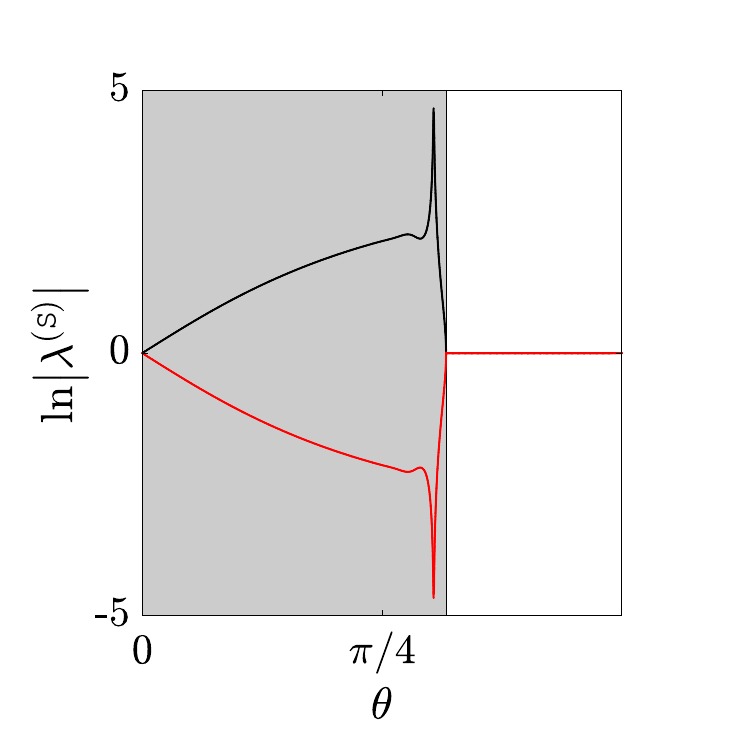}}

\caption{Poles (black circles) and zeroes (red crosses) of the scattering matrix
in the $\left(\protect\phaz/\protect\xr,\protect\ang\right)$ space,
for $\protect\grating/\protect\ksh=1/30$ and $\left(\text{a}\right)$
$\protect\eztep=10^{-2}$, $\left(\text{b}\right)$ $10^{-1}$ and
$\left(\text{c}\right)$ $10^{-1/2}$. Logarithm of the eigenvalues
of $\protect\sr$ as function of $\protect\ang$, for $\protect\eztep=10^{-2}$
{[}panel (d){]}, $10^{-1}$ {[}panel (e){]} and $10^{-1/2}$ {[}panel
(f){]}; we highlight the broken region at which the eigenvalues are
non-unimodular by gray.}

{\small{}{}\label{fig:poleszeroes}}{\small\par}
\end{figure}
The extreme scattering values in the latter case hint at the existence of a pole
of the scattering matrix near $\phaz/\xr=-2/5$ and $\ang=0.09$.
This observation motivates us to evaluate the eigenvalues of the scattering matrix
eigenvalues in the $\left(\ang,\phaz/\xr\right)$ space.  By doing
so, we can identify the poles and zeroes of the scattering matrix,
which we plot in Fig.\ \ref{fig:poleszeroes} using black circles
and red crosses, respectively, for $\eztep=10^{-2}$ $\left[\text{panel }\left(\text{a}\right)\right]$,
$10^{-1}$ $\left[\text{panel }\left(\text{b}\right)\right]$ and
$10^{-1/2}$ $\left[\text{panel }\left(\text{c}\right)\right]$, when
$\grating/\ksh=1/30$. We observe that the poles and zeroes are distributed
symmetrically with respect to the $\pt$-symmetric perturbation $\phaz=0$,
namely, if at a certain $\left(\phaz,\ang\right)$ pair there is a
pole (zero) then at $\left(-\phaz,\ang\right)$ there is a zero (pole). This is in accordance with our analysis in Sec.\ \ref{sec:Scattering-analysis}, where we showed that product of
the modulus of the eigenvalues of $\sr\left(-\phaz\right)$ and $\sr\left(\phaz\right)$
is $1$. 
Such distribution is similar to the symmetric distribution of poles and zeroes
in the complex frequency plane that was studied by \citet{ge2012conservation}.
There, this distribution resulted from the $\pt$ symmetry of the
medium, while in our case, it is a consequence of Eq.\  \eqref{eq:scattering matrix behaviour-1}.

A comparison of the different panels in Fig.\  \ref{fig:poleszeroes} shows that by increasing the amplitude of the modulation, the number
of poles and zeroes increases too. The collection of all these points
constitutes an arc-like structure, whose tip is associated with the
angle at which $\pt$ transition occurs for the $\pt$-symmetric system
defined by $\phaz=0$. To show this, we evaluate the logarithm of
the eigenvalues of $\sr$ when  $\phaz=0$ as function of $\ang$. Specifically, panels (d)-(f) correspond to  $\eztep=10^{-2},10^{-1}$ and $10^{-1/2}$, respectively; we highlight the broken region at which the eigenvalues are
non-unimodular by gray. Indeed, we observe how the tip of the arc-like
structure and the transition angle from gray to white coincide. While we are unable to derive a mathematical explanation for this observation, we note that similar observations were made in Refs.\,\cite{chong2011p,ge2012conservation}, which were later explained in Ref.\,\cite{Ambichl2013prx}. Specifically, figure 8 in Ref.\,\cite{ge2012conservation} shows that at the phase transition of the $\mathsf{S}$ matrix there is an anticrossing of the poles in the complex frequency plane, see also figure 3 in Ref.\,\cite{chong2011p}. There, the symmetric distribution of the poles and zeros about the imaginary axis is a result of the $\mathcal{PT}$ symmetry of the system; here, the $\left(\theta,z_0\right)$ space plays a similar role, where the symmetric distribution of the poles and zeros is about $z_0=0$, owing to Eq.\,\eqref{eq:scattering matrix behaviour-1}. 

Fig.\ \ref{Fig:SrPTPhaseForPowerOfMinusHalf}
also displays a divergence of the eigenvalues, as one of them tends to explode and the other tends to vanish. This is in accordance with the proximity of a zero and a pole near  
$\phaz=0$ in panel (c), where we recall that their overlap corresponds to CPA-laser states. We thus refer to these states, where a zero and pole are very close but not coincide, as a quasi CPA-laser state. In such states, as we highlight next, one of the eigenvalues tends to vanish, but remains greater from zero, while the other grows significantly, but remains finite. This behavior of the eigenvalues implies that the scatterer can simultaneously absorb almost completely incoming coherent waves and significantly enhance certain incoherent excitations. 
\floatsetup[figure]{style=plain,subcapbesideposition=top}

\begin{figure}[t]
\centering\sidesubfloat[]{\label{fig:PoleScEigAgainstPhazWT}\includegraphics[scale=0.33]{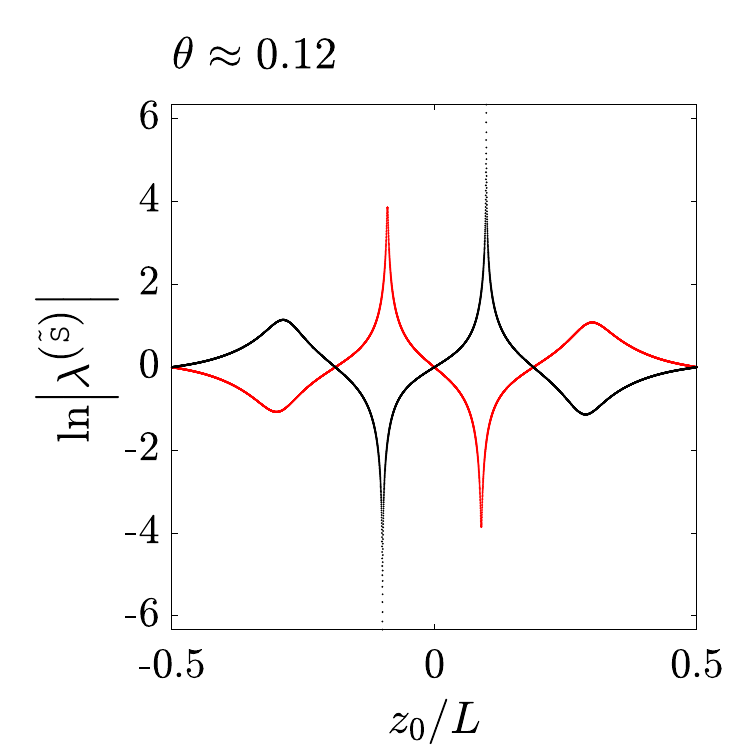}} \quad \centering\sidesubfloat[]{\label{fig:PoleScEigAgainstAngWP}\includegraphics[scale=0.33]{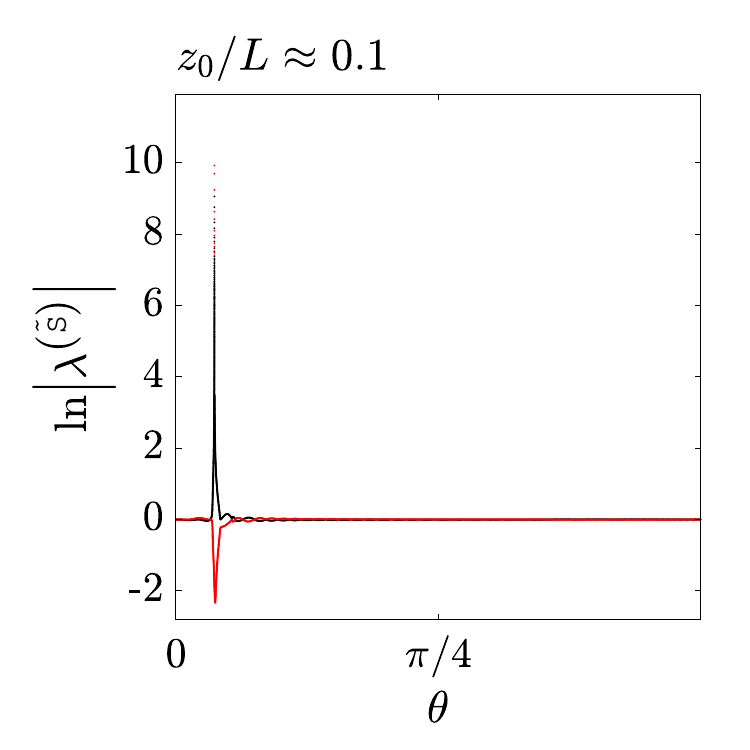}}

\caption{Logarithm of the modulus of the eigenvalues of $\protect\st\left(\protect\eztep=10^{-2}\right)$
$\left(\text{a}\right)$ as function of the phase with fixed incident
angle, $\protect\ang\approx0.12$, $\left(\text{b}\right)$ as function
of the incident angle with fixed phase, $\protect\phaz\approx0.1$.}

{\small{}{}\label{fig:WPWT}}{\small\par}
\end{figure}

Quasi CPA-laser states occur also, e.g., for $\eztep=10^{-2}$ at $\ang\approx0.12$, when $\phaz/\xr\approx\pm0.1$. We examine in Fig.\ \ref{fig:WPWT} how the logarithm of the eigenvalues modulus of  $\sr$ vary near these states as function of two different variables:  (\emph{i}) $\phaz/\xr$ when theta is fixed to $\approx0.12$ {[}panel (a){]}; and  (\emph{ii}) $\ang$ when $\phaz/\xr$ is fixed to  $\phaz\approx0.1$  {[}panel (b){]}, while the rest of the parameters are set to $\grating/\ksh=1/30$
and $\eztep=10^{-2}$. We observe in panel (a)  the symmetric distribution of each one of the eigenvalues with respect to $\phaz=0$. In addition, we observe how near $\phaz/\xr\approx-0.1$ one of them grows and the other diminishes, where near $\phaz/\xr\approx0.1$ their tendency is interchanged. The rapid growth and decay of the eigenvalues near the critical angle $\ang\approx0.12$ is demonstrated in panel (b).     

\floatsetup[figure]{style=plain,subcapbesideposition=top}

\section{Summary\label{sec:Summary}}

We derived an exact solution to the problem of monochromatic oblique
TE waves that are scattered by a waveguide with sinusoid gain-loss
modulation. We have analyzed a family of modulations that are parametrized
by their phase, amplitude and wavelength. We have investigated how
these parameters, together with the frequency and incident angle of
the waves, affect the emergence of lasing, perfect absorption, and
anisotropic transmission resonances. For two particular modulations
that satisfy $\pt$ symmetry, we have evaluated the diagram of the
exact- and broken $\pt$ phases in the parameter space, and the phase
diagram that defines the anisotropic transmission resonances. The two modulations have different diagrams, one of which is much richer, exhibiting multiple re-entries to the broken region. With regard to the diagrams of the anisotropic transmission resonances, we have showed that there is a regime in the parameter space where these resonances
reside very close to Fabry-P\'erot resonances. A further investigation of this property has 
provided guidelines on how to design bidirectional
zero-reflection states. Finally, we have analyzed the poles and
zeros of the scattering matrix, which correspond, respectively, to
lasing and perfect absorption states. We showed that the the modulation
phase and incident angle constitute a design space for these singularities
when the scatterer is subjected to monochromatic waves. Specifically, we have identified quasi CPA-laser states, when a zero and pole are very close but not coincide. These states correspond to scatterers that can simultaneously absorb almost completely incoming coherent waves and significantly enhance certain incoherent excitations.

\section*{Acknowledgments }

We thank anonymous reviewers for constructive feedback that helped us improve this paper. This research has received funding from the European Research Council (ERC) under the European Union's Horizon 2020 Research and Innovation Programme, grant agreement no. 101045494 (EXCEPTIONAL), and the Israel Science Foundation, Israel
Academy of Sciences and Humanities (Grant no.$\ $2061/20).

\appendix

\global\long\def\theequation{A.\arabic{equation}}%
 \setcounter{equation}{0}

\section*{\label{Appendix-A}Appendix. Approximated solution near the first Fabry-P\'erot resonance}

\citet{lin2011prl} have showed that for normal incident waves 
$\left(\ang=\pi/2\right)$ impinging on a $\mathcal{PT}$-symmetric Bragg scatterer, the scattering matrix $\st$ exhibits
an EP at the Bragg point. Here, using a multiple scale expansion, we
show a similar result for oblique waves, impinging on a non-Hermitian scatterer exhibiting one period of modulation, where the EP occur at the  first Fabry-P\'erot resonance, i.e., $\grating=\ksh\sin\ang$. To show this, we assume that
the detuning $\kmb\coloneqq\grating-\ksh\sin\ang$ is very small relatively
to the perturbation wavenumber. We further assume that the amplitude
of the perturbation is small, such $\epsa\coloneqq\eztep/\sin^{2}\ang\ll1$.
With these assumptions at hand, we introduce the variables $\scaxo\coloneqq\xl$
and $\scaxt\coloneqq\epsa\xl$, and employ a power expansion of $\Ef$
in the form

\begin{equation}
\Ef=\Efone+\epsa\Eftwo+...,\label{eq:pert theory-1}
\end{equation}
to solve Eq.\ \eqref{eq:zeq2}. Upon substituting Eq.\ \eqref{eq:pert theory-1}
into Eq.\ \eqref{eq:zeq2}, using the transformation $\px{\xl}=\dxo+\epsa\dxt$,
we obtain two equation

\begin{align}
\dxo^{2}\Efone+\beta^{2}\Efone= & 0,\label{eq:order zero-1}\\
\dxo^{2}\Eftwo+\beta^{2}\Eftwo= & \left(\frac{\beta^{2}-\left(\kinx\right)^{2}}{\epsa}-\ksh^{2}\frac{\xi}{\epsa}e^{2i\beta(\xl+\phaz)}\right)\Efone-2\dxo\dxt\Efone,\label{eq:ord1}
\end{align}
for the orders $\mathcal{O}(1)$ and $\mathcal{O}(\epsa)$ respectively.
The solution of Eq.\ \eqref{eq:order zero-1} for the leading order
is
\[
\Efone=\epf(\scaxt)e^{i\beta\scaxo}+\epb(\scaxt)e^{-i\beta\scaxo},
\]
and to determine $\epf$ and $\epb$, we require that the secular
terms in Eq.\ \eqref{eq:ord1} vanish \citep{watts2012introduction}.
This provides

\begin{equation}
\dxo\left(\begin{array}{c}
\epf\\
\epb
\end{array}\right)=\left(\begin{array}{cc}
\frac{\kmb}{i} & \ksh^{2}\frac{\xi}{-2i\kinx}e^{2i\beta\phaz}\\
0 & -\frac{\kmb}{i}
\end{array}\right)\left(\begin{array}{c}
\epf\\
\epb
\end{array}\right).\label{eq:first order ode for the slowly varying amplitudes-1}
\end{equation}
The solution of Eq.\ \eqref{eq:first order ode for the slowly varying amplitudes-1}
is 
\begin{align}
\left(\begin{array}{c}
\epf\\
\epb
\end{array}\right) & =\left[\cos(\kmb\xl)\left(\begin{array}{cc}
1 & 0\\
0 & 1
\end{array}\right)+i\frac{\sin(\kmb\xl)}{\kmb}\left(\begin{array}{cc}
-\kmb & \ksh^{2}\frac{\xi}{2\kinx}e^{2i\beta\phaz}\\
0 & \kmb
\end{array}\right)\right]\left(\begin{array}{c}
\epf(\xls)\\
\epb(\xls)
\end{array}\right).\label{eq:equivalent to 23-1}
\end{align}
Following the standard procedure to calculate $T$ and $\RLorR$,
we obtain

\begin{align}
T= & 1,R_{R}=\frac{\ksh^{4}\frac{\xi^{2}}{4\left(\kinx\right)^{2}}}{|\kmb\cot(\kmb\xr)|^{2}+\kmb^{2}},R_{L}=0,\label{eq:apptrr}
\end{align}
which indeed implies that under the foregoing assumptions, there is
unidirectional reflection that is associated with an EP of $\st$.

\bibliographystyle{plainnat}
\bibliography{bibtexfiletot}

\end{document}